\documentclass[fleqn,usenatbib]{mnras}
\usepackage{xcolor}

\usepackage[T1]{fontenc}
\usepackage{ae,aecompl}
\usepackage{mathptmx}              

\usepackage{amssymb,amsmath} 
\usepackage{graphics,graphicx} 
\usepackage{soul} 

\newcommand{\secref}[1]{\S\ref{#1}}
\def\l{\bigg{(}} \def\r{\bigg{)}} \def\la{\langle} \def\ra{\rangle}
\def\d{\mathrm{d}} \def\v{\mathbf{v}} \def\B{\mathbf{B}}

\def\nui{{\nu_i}}
\def\msun{M_\odot}

\newcommand{\pder}[2]{\frac{\partial #1}{\partial #2}}

\title[HMNS Disk Outflows in 3D-MHD]{Secular Outflows from 3D-MHD Hypermassive Neutron Star Accretion Disk Systems}

\author[Fahlman, Fern\'andez, \& Morsink]{
Steven Fahlman$^{1}$\thanks{E-mail: sfahlman@ualberta.ca}, Rodrigo
Fern\'andez$^{1}$\thanks{E-mail: rafernan@ualberta.ca}, Sharon
Morsink$^{1}$\thanks{E-mail: morsink@ualberta.ca}\\
$^{1}$Department of Physics, University of Alberta, Edmonton, AB T6G 2E1, Canada\\
}
\pubyear{2023}

\begin{document}
\label{firstpage}
\pagerange{\pageref{firstpage}--\pageref{lastpage}}
\maketitle

\begin{abstract}
Magnetized hypermassive neutron stars (HMNSs) have been proposed as a way for neutron star (NS) mergers to produce high electron fraction, high velocity ejecta, as required by kilonova models to explain the observed light curve of GW170817. The HMNS drives outflows through
neutrino energy deposition and mechanical oscillations, and raises the electron fraction of outflows through neutrino interactions before collapsing to a black hole (BH). Here we perform 3D numerical simulations of HMNS-torus systems in ideal magnetohydrodynamics, using a leakage/absorption scheme for neutrino transport, the nuclear APR equation of state, and Newtonian self-gravity, with a pseudo-Newtonian potential added after BH formation. 
Due to the uncertainty in the HMNS collapse time, we choose
two different parameterized times to induce collapse.
We also explore two initial magnetic field geometries in the torus,
and evolve the systems until the outflows diminish significantly 
($\sim 1 - 2\, \mathrm{s}$). We find bluer, faster outflows as compared to equivalent BH-torus systems, producing $M\sim 10^{-3} M_\odot$ of ejecta with $Y_e \geq 0.25$ and $v \geq 0.25c$ by the simulation end.  
Approximately half the outflows are launched in disk winds at times $t\lesssim 500 \, \mathrm{ms}$, with a broad
distribution of electron fractions and velocities, depending on the initial condition. The remaining outflows are thermally-driven, characterized by lower velocities and electron fractions. Nucleosynthesis with tracer particles shows patterns resembling
solar abundances in all models. 
Although outflows from our simulations do not match those inferred from two-component modelling of the GW170817 kilonova, self-consistent multidimensional detailed kilonova models are required to determine if our outflows can power the blue kilonova.
\end{abstract}

\begin{keywords}
accretion, accretion disks -- MHD -- neutrinos -- nuclear reactions, nucleosynthesis, abundances 
           --- stars: black holes --- stars:neutron
\end{keywords}

\section{Introduction} \label{sec:intro}

Neutron star (NS) mergers are long thought to be a site for the creation of r-process
elements and the production of kilonovae, powered by their radioactive decays
\citep{Lattimer1974, Paczynski1986, Eichler1989,LP1998,Metzger2010}. This was confirmed following the
coincident gravitational wave and kilonova detection, GW170817
\citep{ligo_gw170817_gw,ligogw170817multi-messenger}, which indicated broad agreement
between many observations and theoretical predictions (e.g., \citealt{Cowperthwaite2017,chornock2017,
drout2017,Tanaka2017,tanvir2017}). 

However, many details of the
kilonova and r-process production have not yet been resolved. One of the
outstanding issues comes from multi-component light curve fits using analytic ejecta models to kilonova observations 
that predict a fast ($v\gtrsim 0.25c$), optical ``blue" component composed of high electron
fraction ($Y_e$) material, as well as a slower $v\sim0.1c$ infrared ``red"
component containing most of the ejecta mass with $Y_e \le
0.25$ (e.g., \citealt{Kasen2017, Villar2017,Perego2017}). 
While these multi-component models are simple to apply, multi-dimensional radiative transfer models including accurate nuclear heating rates, thermalization efficiencies, and opacities are more physically motivated. Some of these more advanced kilonova models indicate that in order to recreate the kilonova of GW170817, the disk wind may may require velocities as low as $v\sim 0.05c$ (e.g., \citealt{Ristic2023,Bulla2023}).
3D magnetohydrodynamic (MHD) numerical simulations of black hole (BH) accretion disks from NS mergers, have difficulty producing a massive, high velocity blue component, making the need apparent for more detailed kilonova models, an additional component in NS merger simulations, or both (e.g., \citealt{Metzger2018,Kawaguchi2019,Darbha2020,Barnes2021,Heinzel2021,Korobkin2021,Kawaguchi2022,Ristic2023,Bulla2023,Just2023,Kawaguchi2023}).

Due to differences in requisite physics and computational costs, many 3D numerical simulations split the evolution of NS mergers into two or more phases, usually modelling either the inspiral and merger of the two stars, often in GRHD 
(e.g., see \citealt{Radice2020Review,Perego2021Review, Rosswog2022Review,Janka2022Review} for recent reviews), or the post-merger phase consisting of a torus around a remnant compact object in (GR)MHD (a BH or NS e.g.,
\citealt{HosseinNouri2017,Siegel2018,Fernandez2019,Christie2019,Miller2019,Just2021,FahlmanFernandez2022,
Curtis2023}), although many recent works deviate from this structure (see e.g.,
\citealt{Ciolfi2019,Hayashi2021,LopezArmengol2022, Just2023, Kiuchi2022}). In equal mass neutron star mergers, the secular wind stage is thought to provide the majority of the mass outflow responsible for powering the kilonova, making simulations of this phase crucial for resolving discrepancies with kilonova models of GW170817 (e.g., \citealt{Wu2016,Radice2018Ensemble,Metzger2018,Margalit2019, Fujibayashi2020}). 

The above simulations of BH torus systems suggest that winds from the disk can eject sufficient mass to power the red kilonova, but are lacking enough mass in high electron fraction ( $Y_e \gtrsim 0.25)$, high velocity ($v \gtrsim 0.25c$) outflows to create the blue kilonova.  Several studies point to the formation of a short lived hypermassive NS (HMNS) as a resolution to the lack of this ejecta (e.g., \citealt{Metzger2018,FahlmanFernandez2022,Combi2023Jet,Curtis2023HMNS}). A HMNS formed in a merger is supported against collapse to a BH through additional pressure provided by differential rotation and thermal effects. As the differential rotation is removed through angular momentum transport, and thermal energy is lost to neutrinos (e.g., \citealt{Duez2006,Kaplan2014,Hanauske2017,Ciolfi2019,Bernuzzi2020,Janka2022Review} and references therein), the HMNS will collapse on a timescsale of $t \sim 1-100 \,\mathrm{ms}$. The results of previous 3D simulations and 2D viscous hydrodynamic simulations indicate that during the time it is present, the HMNS will 
drive high electron fraction winds via neutrino absorption, magnetic effects, and mechanically driven oscillations. 

A complete model of the HMNS and torus system requires inclusion of general relativity (GR), magnetohydrodynamics (MHD), neutrino transport, a realistic nuclear equation of state (EOS), and timescales long enough to capture the ejecta from both the early HMNS and late-time torus winds ($\sim 1\;\mathrm{s}$). In particular, the HMNS phase of the merger requires very high spatial resolutions in 3D to resolve the wavelength of the most unstable mode of the magnetorotational instability (MRI) within the HMNS and torus, and ensure the MRI dynamo is not suppressed through axisymmetry \citep{Cowling1933}.

In practice, detailed long-term modelling of these effects are eschewed in the name of computational costs, and instead have been mostly simulated in reduced dimensionality in combination with parameterized angular momentum transport, (e.g., \citealt{Lippuner2017,FahlmanFernandez2018,Shibata2021,Fujibayashi2023,Just2023}), with a parameterized dynamo \citep{Shibata2021MeanField}, or in 3D viscous simulations \citep{Perego2014,Foucart2020MC,Nedora2021}

Currently, there are few simulations of 3D MHD HMNS systems, (e.g.,
\citealt{Kiuchi2012,Siegel_2014,Ciolfi2020,Mosta2020,deHaas2022,Combi2023Jet,Curtis2023HMNS})
with only one simulation lasting $\sim 1\;\mathrm{s}$ \citep{Kiuchi2022}. The
varying input neutrino physics, HMNS lifetimes, and magnetic fields strengths
and orientations result in little consensus on the composition, mass, velocities
and ejection mechanisms for their outflows.

Here we present long-term 3D MHD numerical simulations of HMNS and torus systems, starting from varying idealized initial conditions. We utilize a (pseudo)Newtonian potential \citep{Artemova1996}, the nuclear APR EOS \citep{SCMP2019}, and a leakage/absorption scheme to handle energy deposition and lepton number change from neutrinos, as described in Section~\ref{sec:methods}. 
In Section~\ref{sec:results}
we examine our outflows in the context of powering a kilonova while ignoring effects from relativistic jets, which we are unable to model without full GR. In Section~\ref{sec:comparison} we compare to other works and in Section~\ref{sec:conclusion} we conclude. The Appendix describes updates to our previously used neutrino scheme to handle a HMNS fully contained in the computational domain.

\section{Methods} \label{sec:methods}

\subsection{Numerical MHD}

We use a modified version of the \texttt{FLASH4.5} code to run our MHD
simulations in non-uniform 3D spherical coordinates. A full description (along
with code verification tests) is provided in \citet{FahlmanFernandez2022}, here we
give a brief overview. The code solves the Newtonian
conservation equations for mass, momentum, energy, and lepton number, as well as
the induction equation 
\begin{align}
  \label{eq:FLASHDensity} 
  \pder{\rho}{t} &+ \nabla\cdot{[\rho \v]} = 0 \\
  \label{eq:FLASHMomentum} 
  \pder{(\rho \v)}{t} &+ \nabla\cdot{[\rho(\v \otimes \v)-(\B \otimes \B)]}  + \nabla P = -\rho\nabla\Phi \\ 
  \label{eq:FLASHEnergy}
  \pder{(\rho E)}{t} &+ \nabla\cdot{[\v( \rho E + P)-\B(\v\cdot\B)] } =
-\rho\v\cdot \nabla \Phi_{\rm A} + Q_{\rm net}\\
  \label{eq:FLASHLepton}
  \pder{(\rho Y_e)}{t} &+ \nabla\cdot[\rho Y_e \v] = \Gamma_{\rm net},\\
  \label{eq:Induction}
  \pder{\B}{t} &+ \nabla\cdot{[\v\otimes\B - \B\otimes\v]} = 0. 
\end{align} 
where $\rho$ is the density, $\v$ is the velocity, $\mathbf{B}$ is the magnetic field 
(including a normalization factor $\sqrt{4\pi}$) , $Y_e$ is the electron
fraction, and $E$ is the total specific energy of the fluid
\begin{equation}
E = \frac{1}{2}\left(\mathbf{v}\cdot\mathbf{v} +
\mathbf{B}\cdot\mathbf{B}\right) + e_{\rm int}.
\end{equation}
We denote $e_{\rm int}$ the specific internal energy, 
and $P$ is the sum of gas and magnetic pressure 
\begin{align} 
  P &= P_{\rm gas} + \frac{1}{2} \B \cdot \B.
\end{align}  
We use the constrained transport method \citep{Evans1988} to preserve the
solenoidal condition ($\nabla \cdot \mathbf{B} = 0$) while evolving the
induction equation.  We include source terms from the self-gravitating potential
of the fluid, $\Phi$, neutrino heating and cooling,
$Q_\mathrm{net}$, and lepton number change, $\Gamma_\mathrm{net}$.
Our self-gravity scheme utilizes the Newtonian multipole solver of
\citet{Mueller1995}, as implemented in \citet{Fernandez2019WD}. After the HMNS
collapses into a BH, we add a point source term with
the mass of the remnant to the zeroth moment of the multipole expansion using the Artemova potential
\citep{Artemova1996}. We close the equations with the hot APR EOS of
\citet{SCMP2019}. The additional source terms from our neutrino scheme
are calculated using a 3 species, two source, lightbulb-leakage scheme
\citep{MF14,Lippuner2017,FernandezRichers2022}. We use local emission rates from 
\citet{Bruenn1985}
 for neutrino production from charged-current interactions, 
 and the rates from \citet{Ruffert1997} for
plasmon decay, nucleon-nucleon brehmsstrahlung,
and pair processes. We take into account absorption due to
charged-current weak interactions from electron-type neutrinos and antineutrinos emitted by the central
HMNS and by a ring within the torus. The implementations of our neutrino source terms are detailed in Appendix~\ref{sec:AppA}.

\subsection{Initial Torus}

We initialize our simulations with a constant specific angular momentum,
entropy, and composition torus in hydrostatic equilibrium with a point mass
remnant in a Newtonian potential. Generating the torus requires choosing input
parameters, which we set according to inferred values from GW170817 (See e.g., \citealt{LVSC2017a, Shibata2017,FahlmanFernandez2018} ). For all
our simulations, we set a remnant mass of $2.65 \msun$, torus mass of $0.1
\msun$, radius of density maximum at $50 \,\mathrm{km}$, entropy of $s = 8\,
k_b/\mathrm{baryon}$ and $Y_e = 0.1$. This results in a torus with an initial
maximum density $\rho_{\rm max} = 8\times 10^{10} \,\mathrm{g\,cm}^{-3}$. Matter initialized in the torus is flagged as torus material using a passive mass scalar which is advected with the flow. We neglect the
self-gravity of the torus in the initialization process, as the system mass is
dominated by the remnant. 
We additionally neglect spatial variations in the composition and entropy, which
have been found to make small ($\sim 10\%$) differences in the outflows around BH accretion
disks \citep{Fernandez2017,Nedora2021,Fujibayashi2020Disks,Fujibayashi2020,Most2021}.

The torus is then threaded with a magnetic field. We choose magnetic field configurations that cover possible geometries found in magnetized merger simulations, which generally yield a combination of (turbulent) toroidal and poloidal fields. Due to the high resolution requirements to capture the mechanisms for amplifying the field at merger and within the remnant plus torus (e.g., the Kelvin-Helmholtz instability and MRI), we start with field strengths that assume that growth and saturation has been reached, allowing us to resolve the smallest growing wavelength of the MRI with our grid (See Figure~\ref{fig:mriquality} and Section~\ref{sec:computationaldomain}). These field strengths are similar to those found 
in merger simulations which use sub-grid models to resolve these effects (e.g., \citealt{Aguilera-Miret2023}). For models which specify a poloidal field geometry, this is initialized using an azimuthal vector potential
that follows the density,
\begin{align}
  A_\phi \propto \mathrm{max}(\rho - \rho_0, 0) 
\end{align}
where $\rho_0 = 0.009\rho_\mathrm{max}$ is the cutoff density used to ensure the field
is embedded in the torus. This results in a poloidal (``SANE") field topology
\citep{Hawley2000}. The field is normalized to be dynamically unimportant, with
an average plasma $\beta$ of
\begin{equation} 
  \langle \beta \rangle = \frac{\int P_\mathrm{gas} \mathrm{d}V}{\int
P_\mathrm{mag}\mathrm{d}V} = 100,
\end{equation}
resulting in a maximum field strength of $|B_r| \sim |B_\theta| \sim 4\times
10^{14}\,\mathrm{G}$. Models with a toroidal field geometry are initialized as
\begin{align}
  B_\phi = B_0\mathrm{max}(\rho - \rho_0,0),
\end{align}
where $B_0 = 5 \times 10^9\,\mathrm{G} / (\mathrm{g\, cm}^{-3}) $.
This results in nearly constant azimuthal magnetic field embedded within the
torus, which tapers off as a function of the density profile. The field has a
maximal value of $8 \times 10^{14}\,\mathrm{G}$, and an average strength of
$\sim2\times 10^{14} \, \mathrm{G}$.
 
\subsection{Initial HMNS}
Our torus orbits a stable equilibrium model of an azimuthally-symmetric,
differentially-rotating NS. We find 2D HMNS thermodynamic profiles using the
code \texttt{NewtNeut}, a reduced version of the code \texttt{drns}
\citep{Stergioulas1995}. \texttt{NewtNeut} generates differentially-rotating
high-density stars in the Newtonian limit, taking a central density, rotation
law, and EOS as input, and utilizing the self-consistent field method
of \citet[HSCF]{Hachisu1986} to generate thermodynamic profiles for a
differentially rotating non-relativistic star. The equilibrium equation that
the HSCF method solves is 
\begin{align} \label{eq:HSCF}
  \int \rho^{-1} \d P = C - \Phi_g + \int \Omega^2(r_\mathrm{cyl})
r_\mathrm{cyl} \d r_\mathrm{cyl},
\end{align}
where $\Phi_g$ is the (Newtonian) gravitational potential, and $\Omega$ is a
choice of rotation profile dependent on $r_\mathrm{cyl}$, the cylindrical
radial coordinate. 
We use the well studied $j$-const rotation law
\citep{Hachisu1986,Baumgarte2000} 
\begin{align}
  \Omega = \frac{\Omega_0}{(1 + \hat A ^2 \hat r^2\sin^2\theta)},
\end{align}
where $\Omega_0$ is the central rotation rate, $\hat A = A/r_e$ is a scaling
constant which sets the amount of differential rotation, and $\hat r = r/r_e$ is
the radial coordinate normalized to the equatorial radius of the star. We use a
uniform spherical grid with a resolution of $(N_r\times N_\theta)
=(200\times200)$ to generate one quadrant of the NS, and then assume azimuthal
and equatorial symmetry to generate the rest of the star. The HSCF method
requires finding the quantity $\int \rho^{-1} \d P$, which can be obtained from the
enthalpy,
\begin{align} \label{eq:drnsenth}
  h = \int \rho^{-1} \d P + \int T \d S,
\end{align}
as long as our star satisfies the barotropic condition $T \d S=0$. We choose
to generate stars of constant entropy, so that neutrino emission is not
suppressed by a zero temperature initial condition. Our choice of constant
angular momentum differential rotation law and constant entropy are not entirely
correct for the remnant of a NS-NS merger, as more realistic neutron stars
follow a rotation law that peaks away from the central density, (e.g.,
\citealt{Hanauske2017,Iosif2022}) and have varying spatial entropy (e.g.,
\citealt{Most2021,Nedora2021}).

The generated HMNS has a mass 2.65$\msun$, entropy $s= 2\,k_\mathrm{B}/\mathrm{baryon}$, with $\hat
A=0.5$, and the central rotation rate chosen such that the ratio of polar to
equatorial radius is $r_p/r_e = 0.75$. This makes a star with a maximal
equatorial radius and central rotation rate of 23.5\,km and
$3780\,\mathrm{rad\,s}^{-1}$, respectively, corresponding to a period of $\sim
1.6\,\mathrm{ms}$. The thermodynamically consistent axisymmetric density,
electron fraction, temperature, internal energy, and rotational velocity profile
output from \texttt{NewtNeut} are read into \texttt{FLASH}, and the relevant
quantities are linearly interpolated to the cell centers to create our initial HMNS. Like the torus, HMNS matter is also flagged using a passively advected mass scalar.
In all our runs, we embed the HMNS with a dynamically unimportant axisymmetric
toroidal field that vanishes as we approach the origin, as motivated by
merger simulations \citep{Shibata2021MeanField}. The HMNS field is given by
\begin{align}
  B_\phi = B_{NS} \l \frac{r}{r_e}  \r^2 \mathrm{max}(\rho - \rho_0,0)
\end{align}
where $\rho_0 = 5\times 10^{13}\,\mathrm{g\,cm}^{-3}$ is $1/10$ the central
density of the HMNS, and the constant $B_{NS} = 8\times
10^{4}\,\mathrm{G}/(\mathrm{g\,cm}^{-3})$
is set such that the total magnetic
energy within in the NS is $2\times 10^{47}\,\mathrm{erg}$. We note that the
magnetic field geometry is no longer toroidal by $t \sim 5\,\mathrm{ms}$, as it
is modified by the dynamics inside the HMNS. 

\subsection{Tracer Particles}
Tracer particles are used to track nucleosynthesis in post-processing. We initialize 10,000 tracer particles by placing them into pseudorandom positions within the domain, following the density distribution, where the density falls between $10^6 \leq \rho \leq 10^{13} \,\mathrm{g\,cm}^{-3}$ and the atmospheric fraction is 0. This ensures that the particles are embedded within the torus and the edges of the HMNS, but also that few are trapped in the HMNS when it collapses. Particles are then advected with the fluid flow, and the history of any particles that make it past a fixed extraction radius $r_{\rm out}$ are used in nucleosynthesis calculations. The latter are carried out with the publicly
available nuclear reaction network \texttt{SkyNet} \citep{Lippuner2017Skynet}, with the same settings as \citet{Fernandez2020BHNS}.

\subsection{Computational Domain} \label{sec:computationaldomain}
We solve the MHD equations in spherical polar coordinates $(r,\theta,\phi)$ on a
domain initially ranging from an inner boundary at $r_\mathrm{min}=0\,\mathrm{km}$ to an
outer radius $r_\mathrm{max}=10^5\,\mathrm{km}$, located far away from the
central remnant.  Both the polar $(\theta)$ and azimuthal ($\phi$) domains
subtend an angle from 0 to $\pi$, 
creating a hemisphere. 

Upon collapse of the
HMNS, we excise an integer number of cells covering a radial region interior to a new radius $r_\mathrm{min}$,
located approximately halfway between the event horizon and ISCO of the newly formed BH. 
To preserve the total number of radial cells in the grid, the domain is expanded outwards in radius by the same number of cells removed around the origin, with the new cells filled with ambient medium. Since no ejecta has reached the outer radial boundary at the collapse times, this results in no effective change to the simulation outside $r_{\rm min}$. For all models, this results in a post-collapse value
of $r_\mathrm{min} = 15.4\,\mathrm{km}$ and an outer boundary at
$r_\mathrm{max} = 5.23 \times 10^5\,\mathrm{km}$.

\begin{figure} 
\includegraphics[width=\columnwidth]{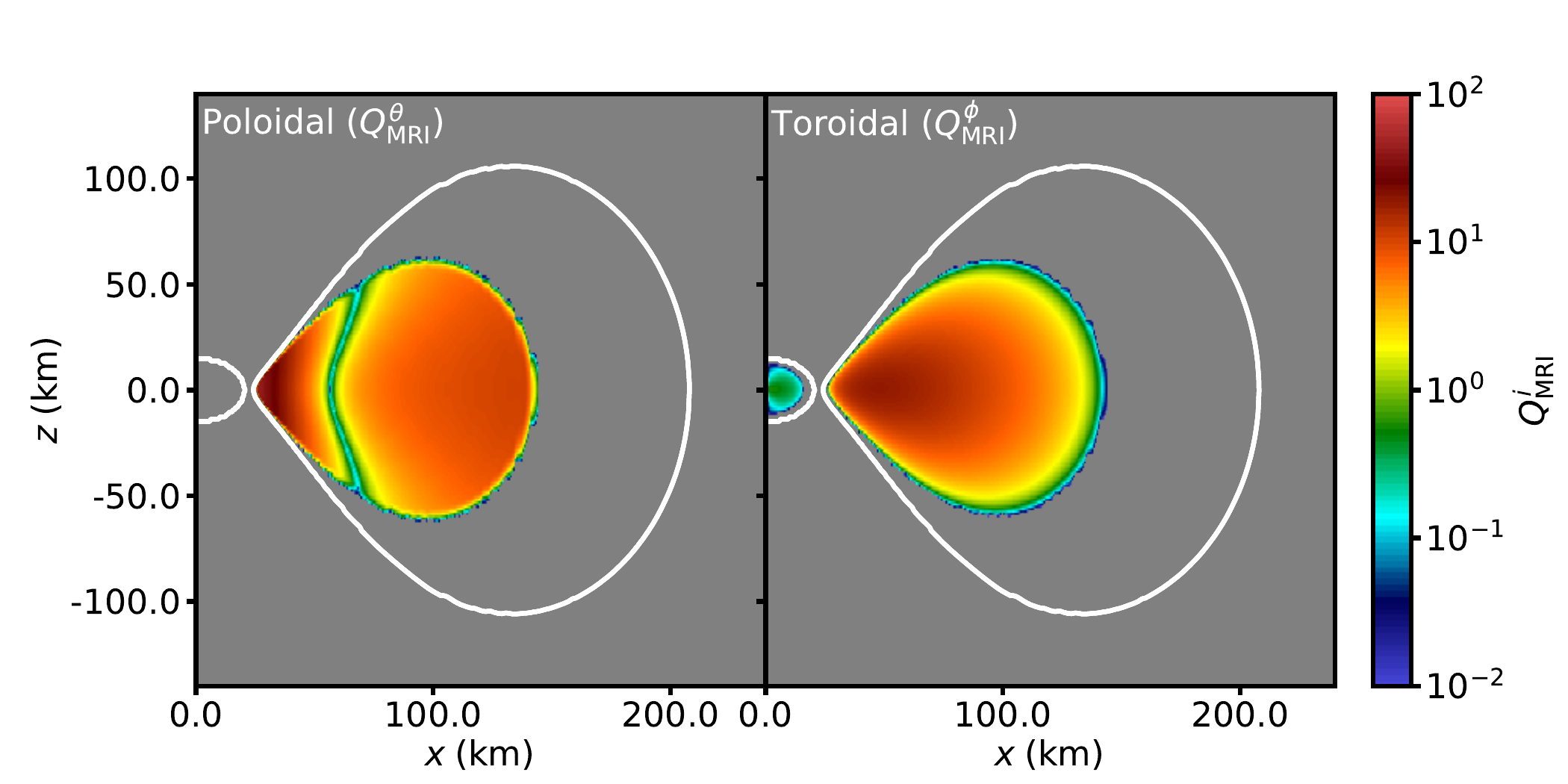}
\caption{Slices of the MRI Quality parameter $(Q_\mathrm{MRI}^i =
\lambda_\mathrm{MRI}^i/\Delta \ell_i )$, where $\lambda_\mathrm{MRI}^i$ is the wavelength of the most
unstable mode of the MRI and $\Delta \ell_i $ is the cell width in the specified $i$-th direction. Slices are in the  $x-z$ ($\phi=0$) plane of the \emph{Bpol-t30} and \emph{Btor-t30} models at initialization. The solid white 
lines mark the $3\times10^8\, \mathrm{g\,cm}^{-3}$ density contour,
corresponding to near the edge of the torus and HMNS. Gray colour
denotes regions without any magnetization in the relevant direction. The field is well embedded within the
torus, and is generally resolved by at least 10 cells. The toroidal field in the
HMNS is not well resolved due to the high densities, and as such we do not
expect to resolve magnetized outflows driven by MRI turbulence in the remnant.
}
\label{fig:mriquality}
\end{figure}

We use a logarithmic grid in the radial direction, a grid evenly spaced in
$\cos\theta$ in the polar direction, and uniform spacing in azimuth. To
avoid issues with time stepping close to the singularity, we make the innermost
radial cell large, with a size of $3\,\mathrm{km}$, encompassing the inner $\sim$13\%
of the HMNS radius. We set our polar and azimuthal resolution such that the
initial wavelength of the most unstable mode of the MRI within the torus is resolved with $\sim$
10 cells, and set a radial resolution to get approximately square cells in the
midplane of the torus. This results in a mesh size of $(N_r \times
N_\theta \times N_\phi) = (580\times120\times64)$, focused towards the midplane
of the disk, with $\Delta r/r \sim 0.018$, $\mathrm{min}(\Delta \theta) \sim
0.017$, and $\Delta \phi \sim 0.1$. We show the MRI quality factor \citep{Sano2004} for both our
initial setups in Figure~\ref{fig:mriquality}. 

The boundary conditions of our domain are periodic in the azimuthal direction, and
reflecting along the polar axes. The outer radial boundary is set to outflow,
while the inner radial boundary is initially reflecting while the HMNS survives.
After collapse, we set the inner radial boundary to outflow so matter can
accrete onto the BH.

\subsection{Floor}
\label{sec:floor}

To avoid issues with convergence, finite volume codes implement density floors.
We impose a spatial and temporal varying floor to prevent computationally
unfeasible low timesteps in magnetized regions near the inner boundary and
rotation axis, while not affecting dynamics of the outflows. We implement the
functional form used in \cite{FahlmanFernandez2022}
\begin{align}
  \rho_\mathrm{floor,A} = \rho_\mathrm{sml} \l \frac{r}{20\,\mathrm{km}} \r^{-3} \l
\frac{1}{\sin\theta} \r^{0.75} \l
\frac{\mathrm{max}[t,0.1\,\mathrm{s}]}{0.1\,\mathrm{s}}
\r^{-1.5},
\end{align}
where $\rho_\mathrm{sml} = 2\times 10^4\,\mathrm{g\,cm}^{-3}$. When the code
reaches a density below the floor value in that cell, we add atmospheric
material to raise it back up. Identical floors are implemented for pressure and
internal energy, with the respective scaling constants $P_\mathrm{sml} = 2\times
10^{14}\,\text{erg\,cm}^{-3}$ and $e_\mathrm{sml} = 2\times
10^{11}$\,erg\,g$^{-1}$. In the magnetized polar funnel close to the black hole,
we find that this restriction is often too low, but raising the floor results in
unphysical effects in important areas of the flow. To address this, we impose an
alternative floor on the density only that is dependent on the magnetization, as in
\citet{Fernandez2019}
\begin{align}
  \rho_\mathrm{floor,B} = 
    \begin{cases}
    \frac{P_\mathrm{mag}^2}{\zeta c^2},\, &r < 200\,\mathrm{km} \\
    0 , &r \ge 200\,\mathrm{km} 
    \end{cases}
\end{align}
where we find $\zeta = 2$ a reasonable value to increase the timestep in the
magnetized funnel, without affecting the dynamics in the disk.
We then impose the floor
\begin{align}
  \rho_\mathrm{floor} &= \mathrm{max}[\rho_\mathrm{floor,A},\rho_\mathrm{floor,B}]
\end{align}
\subsection{Models} 
\label{sec:models}
Our models are summarized in Table~\ref{tab:models}. We choose parameters for
the torus and HMNS which are the most likely properties of the remnant + disk
system for GW170817 (e.g., \citealt{LVSC2017a, Shibata2017,
FahlmanFernandez2018}), while varying the lifetime of the remnant
and the initial field topology. 
 
\begin{table}
\caption{Simulation Parameters. From left to right, they list the name of the
model, the mass of the remnant, the lifetime of the HMNS, the initial mass of
the torus, the peak magnetic field strength in the torus, and the magnetic field
geometry.}
\label{tab:models} 
\resizebox{\columnwidth}{!}{%
\begin{tabular}{lccccc} 
\hline
Model & $M_\mathrm{remnant}$ & $ \tau_\mathrm{HMNS}$&  $M_{\rm t}$ & $ || \mathbf{B} ||$ & $\mathbf{B}$ \\ 
      & ($M_\odot$)       &  (ms)                &  $(M_\odot)$ & (G)           & geom  \\
\hline
\emph{Bpol-t30 } & 2.65 & 30  & 0.10 & $4\times10^{14}$ & pol \\ 
\emph{Bpol-t100} &      & 100 &      &                  &     \\ 
\emph{Btor-t30 } &      & 30  &      &                  & tor \\ 
\emph{Btor-t100} &      & 100 &      &                  &     \\
\hline
\end{tabular}%
}
\begin{flushleft}
\end{flushleft}
\end{table} 

\subsection{Outflows} \label{sec:outflows}

We calculate the total outflow by temporally integrating the mass flux passing
through the an extraction radius $r_\mathrm{out}$,
 \begin{align}
\label{eq:Mtot_simulation}
  M_\mathrm{out} =  \int_t \iint_{A_r} (\rho v_r  \d A_r) \d t, 
\end{align}
where $A_r = \iint r^2\sin\theta\d\theta\d\phi$ is the area of the cell face.
After testing various extraction radii, we choose $r_\mathrm{out} = 1000\,\mathrm{km}$ as a radius where unbound matter has minimal interaction with the atmosphere, which can impact the energy of the ejecta, as well as being  far away from the edges of the viscously spreading disk at late times, but close enough for most ejecta to cross during the simulation time. 
Matter is considered unbound if it has a positive Bernoulli parameter at the extraction radius
\begin{align} 
  \Phi_g + e_\mathrm{int} + e_\mathrm{k} + e_\mathrm{mag}
+\frac{P_\mathrm{gas}}{\rho} > 0,
\end{align}
and we discount any ``atmospheric" matter that is present due to the requirement of the non-zero
floor (\S\ref{sec:floor}).

We tabulate the ejecta in terms of total, ``blue" ($Y_e \geq 0.25)$, and ``red"
($Y_e < 0.25$) outflows, based on the characteristic division between lanthanide-poor
and lanthanide-rich material found in parametric nuclear network
calculations (e.g.,
\citealt{Lippuner2015,Kasen2015}).
The mass weighted averages of electron fraction and radial velocity,
\begin{align} 
\label{eq:Ye_ave}
  \la Y_e \ra = \dfrac{\int_t \iint_{A_r} (\rho v_r Y_e \d A_r) \d t, }{M_\mathrm{out}} \\ 
\label{eq:vr_ave}
  \la v_r \ra = \dfrac{\int_t \iint_{A_r} (\rho v_r v_r \d A_r) \d t, }{M_\mathrm{out}}
\end{align} 
are shown alongside the mass outflows in Table~\ref{tab:outflowresults}. 

Throughout this paper, we identify material that is ejected through different mechanisms using the energy and entropy of the ejecta. Regions with large neutrino heating source terms are imparted with large internal energies and entropies, which we identify as neutrino driven winds. Conversely, fast, low-entropy material from regions with small neutrino heating source terms and high magnetization are identified as purely MHD driven. 
Passive tracer particles are also used to track the source terms applied to them as they are ejected and corroborate the identification of different mass ejection mechanisms.

\section{Results} \label{sec:results}
\subsection{Overview of HMNS Outflows}

\begin{figure} 
\includegraphics[width=\columnwidth]{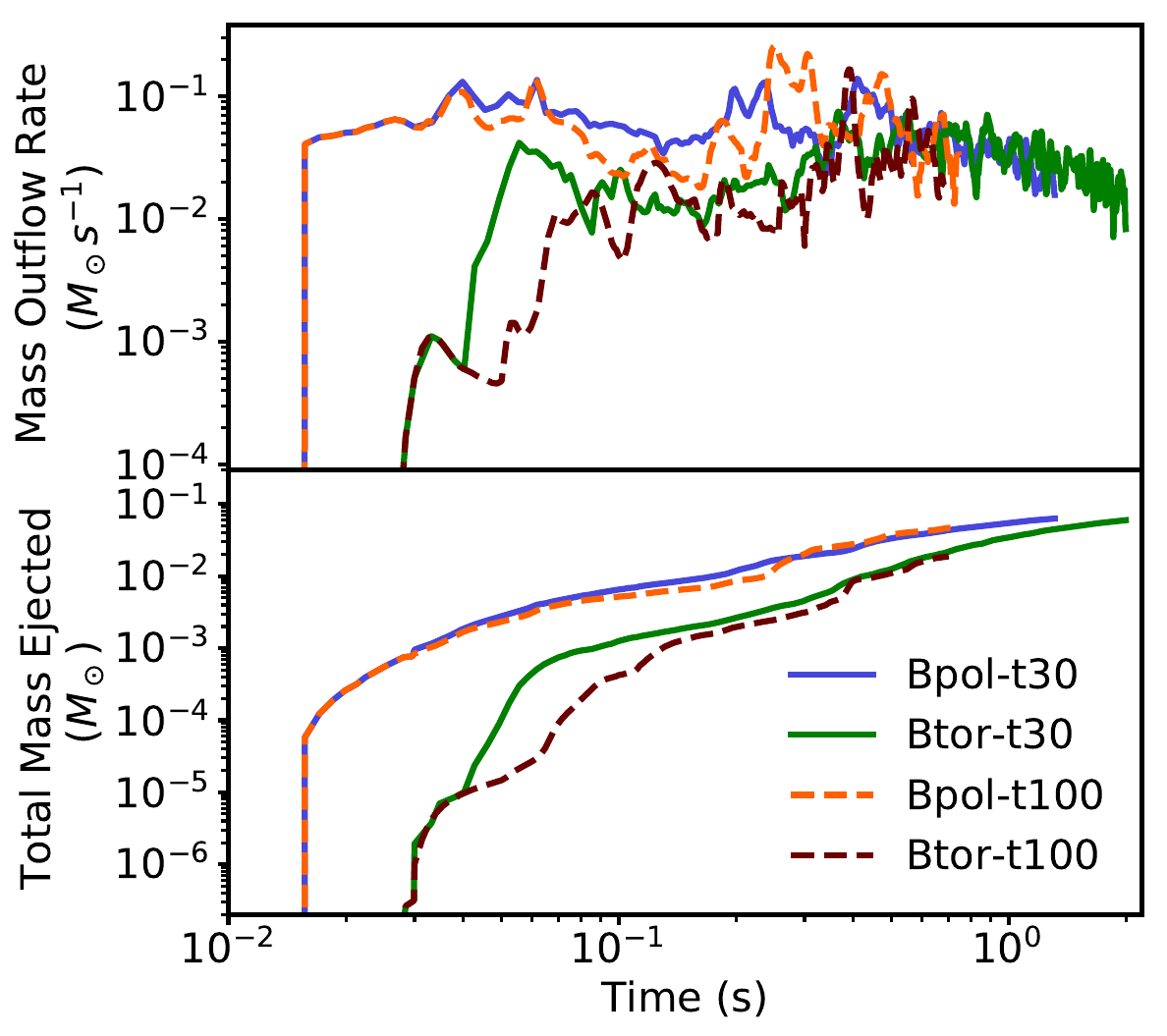}
\caption{\emph{Top:} Unbound mass outflow rates in our simulations at a radius of $r=1000\,\mathrm{km}$. \emph{Bottom:} Cumulative ejected mass in the simulations. } 
\label{fig:massoutflowrates} 
\end{figure} 

We show the mass outflow rates and cumulative mass outflows for all models in Figure~\ref{fig:massoutflowrates}. Mass ejection is dominated by the torus ($\gtrsim 99\%$), although the HMNS provides additional channels.
During the first $\sim 30\,\mathrm{ms}$ of evolution, the HMNS ejects mass through mechanical
oscillations, which  
induces noticeable pressure waves in the torus. Viscous spreading from the action of the MRI and the HMNS oscillations drives the center of mass of the torus out from $\langle r_\mathrm{CM} \rangle \sim 10 \mathrm{km}$ by a factor of 2-4, depending on the initial field geometry and lifetime of the HMNS, and the lower density edges of the torus develop turbulent structures. 

\begin{figure*} 
\includegraphics[width=\textwidth]{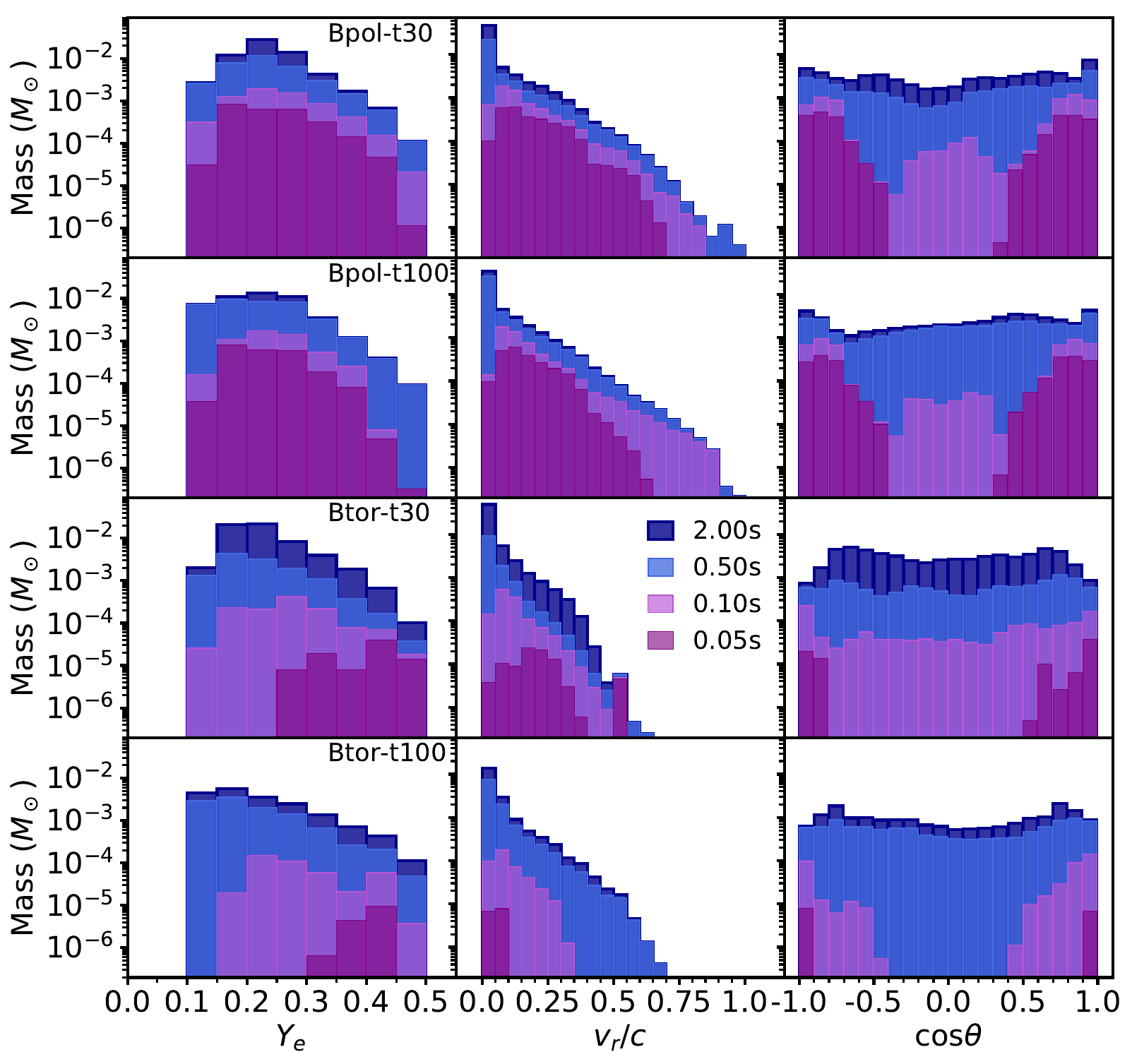}
\caption{Total unbound mass ejected by our simulations at the extraction radius, binned into electron fraction
$Y_e$,radial velocity $v_r$, and polar angle $\theta$. The bin sizes are $\Delta
Y_e = 0.05$, $\Delta v_r/c = 0.05$, and $\Delta \cos\theta = 0.1$, where
$\cos\theta = 0$ is the midplane. Accounting for the time taken for the ejecta to reach the extraction radius, ($r_\mathrm{out}/v_\mathrm{r} \lesssim 70\,\mathrm{ms}$), the 0.1\,s, 0.5\,s and 2.0\,s bins are populated by ejecta launched post-collapse in the \emph{Bpol-t30} and \emph{Btor-t30} models. Similarly, the 0.5\,s and 2.0\,s bins are comprised of mainly post-collapse ejecta in the \emph{Bpol-t100} and \emph{Btor-t100} models.
}
\label{fig:histograms} 
\end{figure*}

The initial sharp increase in mass outflow rate shown in Figure~\ref{fig:massoutflowrates} is dominated by ejection through MHD stresses in the torus that are dependent on the initial magnetic field geometry, as seen in BH-tori simulations \citep{Christie2019,Siegel2019,FahlmanFernandez2022,Curtis2023}. The large scale poloidal field generates characteristic ``wings" in the first $\sim 30\,\mathrm{ms}$, whereas the toroidal field takes an additional ($\sim 50\,\mathrm{ms}$) for dynamo action to convert the geometry into large scale poloidal structures, which then drives outflows. Material ejected through MHD stresses tends to have a wide range of electron fractions and velocities, spanning $Y_e \sim 0.1-0.5$, and $v \sim 0.05 - 0.6c$. The majority of MHD-driven and neutrino-driven winds are imprinted in the levels showing $t \lesssim 500\,\mathrm{ms}$ in the cumulative mass histograms of Figure~\ref{fig:histograms}, with the transition to purely thermally driven outflows (\secref{sec:postcollapse}) happening around the $500\,\mathrm{ms}$ mark.

{Also in the first $30-100\,\mathrm{ms}$, depending on the HMNS lifetime,} high-entropy ($s\sim100\,k_\mathrm{B}/\mathrm{baryon}$), magnetically- and neutrino-driven outflows are driven from the edges of the torus, which achieves electron fractions $Y_e\gtrsim 0.3$ and velocities $v\gtrsim 0.5c$. As illustrated by the angular histograms of Figure~\ref{fig:histograms}, this material is preferentially ejected in a cone of opening angle $50^\circ$, centered on the angular momentum axis (the polar regions). It is dominated by matter from the torus, with $\sim 30$ times more ejecta from the torus than from the HMNS. 

\begin{figure} 
\includegraphics[width=\columnwidth]{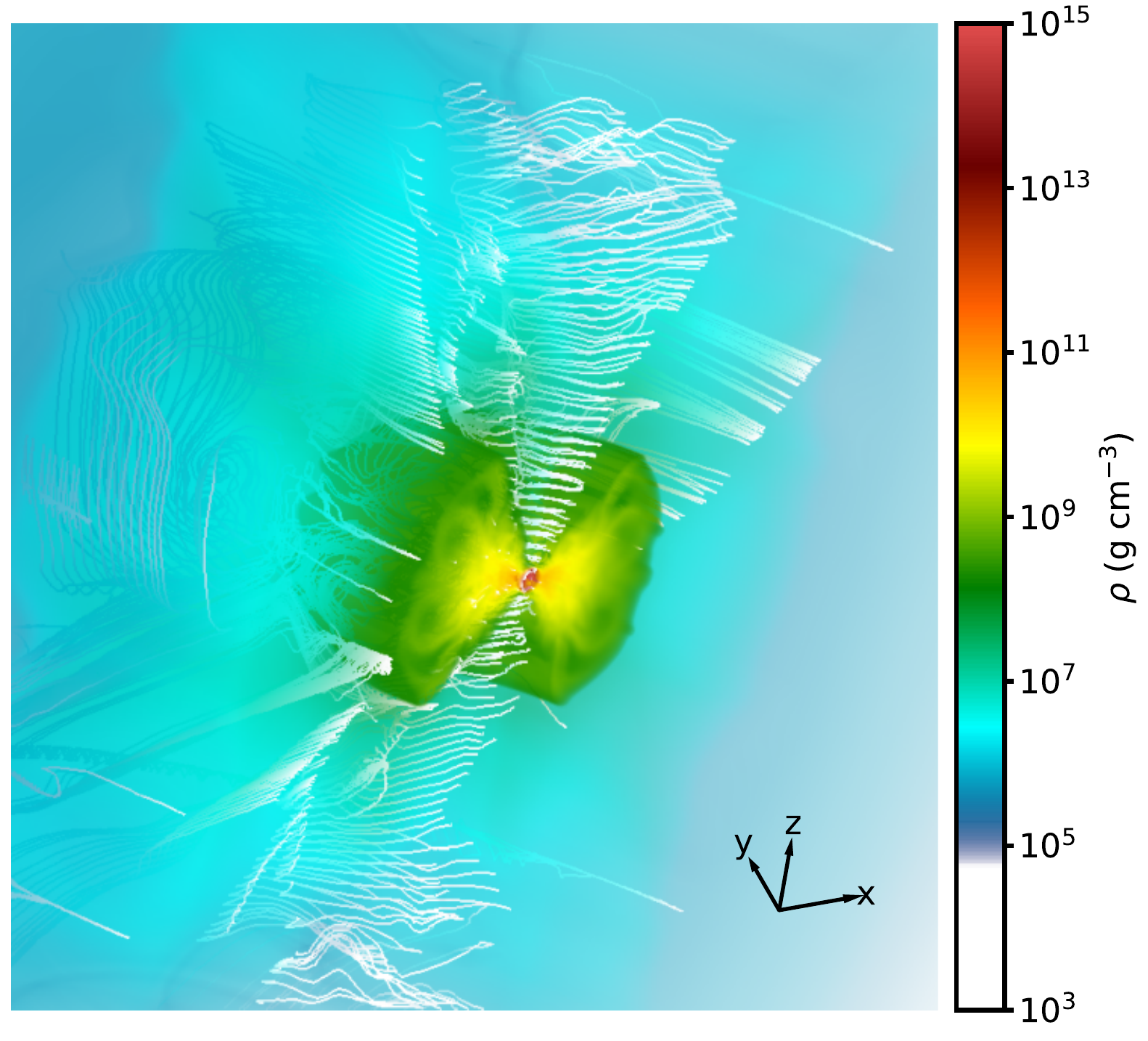}
\caption{Volume rendering of the \textit{Bpol-t30} run at $t=30\,\mathrm{ms}$, right
before the HMNS collapse to a BH is initiated. Colors show the density with
increasing transparency as the density decreases. The visible outer edge of the torus in green ($\sim10^{8}\mathrm{g\,cm}^{-3}$) is located $\sim 350 \,\mathrm{km}$ from the remnant in the midplane. Magnetic field streamlines are shown in white. Note the formation of a lower density, highly magnetized funnel along the spin axis of the HMNS.}
\label{fig:3dplot}
\end{figure}

\begin{figure*} 
\includegraphics[width=\textwidth]{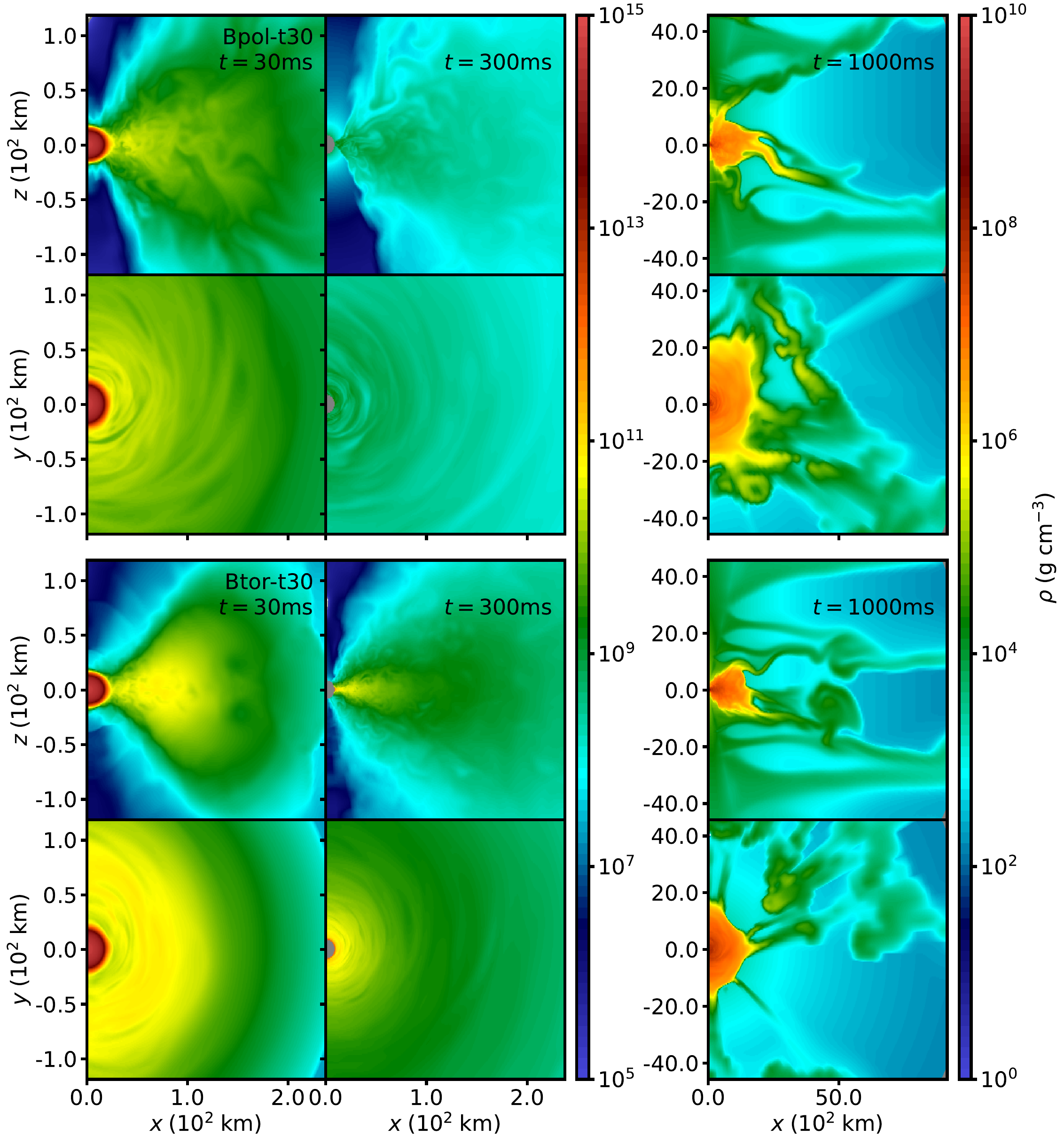}
\caption{Slices of density in the $x-z$ ($\phi=0$, polar) and $x-y$ ($\theta=0$,
equatorial) plane for the models \emph{Bpol-t30} (top two rows) and
\emph{Btor-t30} (bottom two rows) at $t=30\,\mathrm{ms}$ (just before collapse), 300\,ms
and 1\,s. The 2 leftmost columns show the difference in torus structure and
density just before and a few hundred ms after collapse, corresponding to when
most of the outflows occur. Gray regions show the excised black hole, and are
outside the computational domain. Note the changes in both colour and length
scale for the final column, showing the late time periodic mass ejection events
from the torus. 
}
\label{fig:2dsliceplots}
\end{figure*}

Figure~\ref{fig:3dplot} and the leftmost panels of Figure~\ref{fig:2dsliceplots} show the matter density around the HMNS just before collapse. A magnetized funnel is formed while the HMNS survives, but there remains a significant amount of matter in the funnel with densities $\rho \sim 10^6-10^8\,\mathrm{g\,cm}^{-3}$ from HMNS oscillations and accreting matter, leading to the absence of steady-state, high velocity, collimated magnetic ``tower" outflows from the HMNS (the so called ``baryon loading problem"). After collapse of the HMNS, the polar regions become evacuated of matter and sit on our imposed density floor, so we cannot draw conclusions on whether or not matter would be launched post-collapse.

\begin{figure*} 
\includegraphics[width=\textwidth]{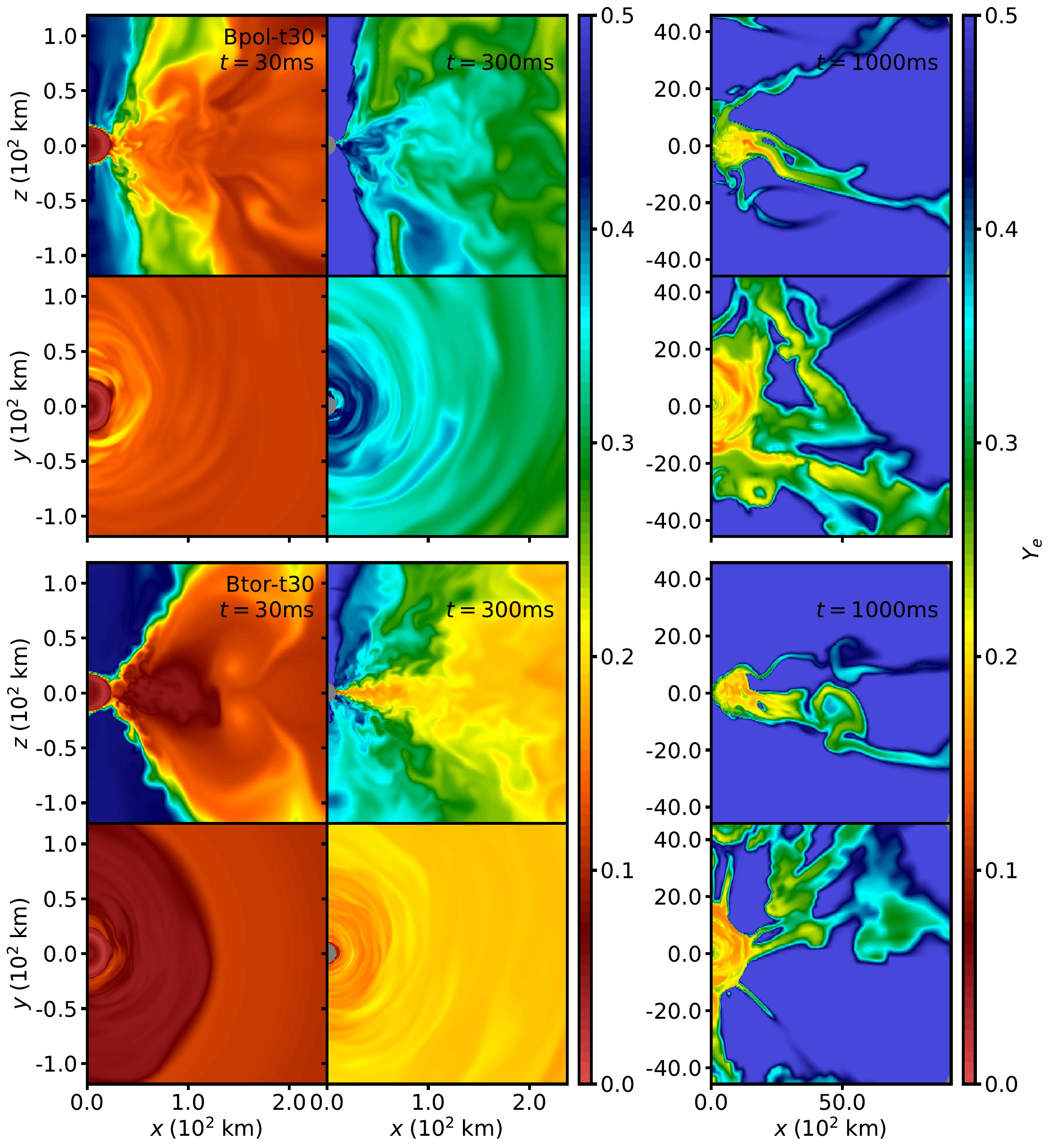}
\caption{Same as Figure~\ref{fig:2dsliceplots}, but for electron fraction.}
\label{fig:2dsliceplots_ye}
\end{figure*}

Shown in Figure~\ref{fig:2dsliceplots_ye} is the spatial distribution of ejecta $Y_e$ for models \emph{Bpol-t30} and \emph{Btor-t30}. The aforementioned magnetized neutrino-driven winds are visible as high $Y_e$ matter in the polar regions, common to all models. The average $Y_e$ of the torus differs between models over the interval $t\sim 10-400$s
from differences in accretion physics during this phase. Due to the susceptibility of the poloidal field to the MRI, the \emph{Bpol} models begin accretion onto the HMNS earlier and at a higher rate than the \emph{Btor} models, and feedback from accretion creates eddies of a size similar to the scale height of the torus. This expansion lowers the average density, thus lowering the attenuation of HMNS irradiation by matter, and mixes irradiated material from the accretion flow back into the dense regions of the torus. This results in an increase in the average electron fraction to $Y_e \sim 0.35$ in the densest midplane of the torus, as compared to $Y_e \sim 0.2$ in the \emph{Btor} runs. 

\subsection{Overview of Post-Collapse Outflows} \label{sec:postcollapse}

After we trigger the collapse of the HMNS, the system then resembles a BH-torus system. In all cases, there is an increase in outflow caused by a magnetized shock wave launched by the collapse of the HMNS, and a subsequent settling into steady accretion onto the newly formed BH. This effect shows up most prominently in the \emph{Btor-t30} model, where it is not masked by other outflows, and is noticeable as a spike in mass ejection at $t\gtrsim 30\,\mathrm{ms}$ in Figure~\ref{fig:massoutflowrates}. In the \emph{Bpol-t30} case, we find that expansion of the torus due to the action of the MRI causes $\sim 10\%$ of the torus mass to not have enough angular momentum to maintain its orbit, plunging into the BH on a timescale of $\lesssim 1\;\mathrm{ms}$ after collapse. For the \emph{Btor-t30} run, we find that only $1\%$ of material is caught in the collapse, in comparison. This is a result of disk spreading and accretion induced by the MRI taking longer to initiate, as well as comparatively weaker initial magnetically driven outflows, resulting in a much more compact torus configuration - the torus centroid is located only $1.1$ times further out than its initial position, as opposed to the $2.2$ times increase in the \emph{Bpol-t30} case. For the longer-lived HMNS cases, \emph{Bpol-t100} and  \emph{Btor-t100}, the longer time for accretion and viscous spreading to occur causes even more mass to be lost upon collapse, $\sim 25\%$ and $\sim 15\%$, respectively, of the torus is accreted instantaneously.

In all cases, material that is ejected by $t \gtrsim 600\,\mathrm{ms}$ is very similar to that of previous BH-torus studies in (GR)MHD, and tends to be in stochastic, slow ($v \lesssim 0.1c$) MRI turbulence driven outflows, with an electron fraction of $Y_e \sim 0.2-0.3$ set by the equilibrium value in the torus (see e.g., \citealt{Siegel2018}). The thermally driven outflows are noticeable as a peak in the velocity and electron fraction histograms in Figure~\ref{fig:histograms} at $v \sim 0.05c$ and $Y_e \sim 0.2$.

The total mass ejected asymptotes to similar values at times $t\sim 1$s in both \emph{Bpol-t30} and \emph{Btor-t30}, although a significantly larger fraction (factor $2$, in comparison) of that mass is contained in redder, slower outflows in \emph{Btor-t30}. While the average electron fraction remains the same in the blue outflows between the two runs, they too tend to be about half as fast in the toroidal model. 

Both \emph{Bpol-t100} and \emph{Btor-t100} are very similar to their $30\,\mathrm{ms}$ counterparts. Although they are run for a shorter amount of simulation time ($t \sim 0.5\,\mathrm{s}$), cumulative mass ejection and mass ejection rates match those from their counterparts well at that time, and we do not expect large differences in the thermally driven outflows past this point in time. This is explained by mass ejection being dominated by MHD stresses in the torus at early times ($t \lesssim 100\,\mathrm{ms}$) and thermal outflows at late times ($t \gtrsim 400\,\mathrm{ms}$), the latter being mostly unaffected by the lifetime of the HMNS. The neutrino-driven winds between 30 and 70\,ms tend to be diminished by the accretion flow onto the HMNS, which cools more effectively than the neutrino irradiation heats it, 
making ejecta in this time from lower density, lower temperature edges of the flow where neutrino heating dominates.  While total outflows remain the same, the proportion of blue outflows when compared at the latest common time ($t=0.5\,s$) is 5\% higher for the the longer lived HMNS models, and they tend to be concentrated in magnetically and neutrino driven winds with velocities $v\sim0.3-0.5c$.

\begin{figure} 
\includegraphics[width=\columnwidth]{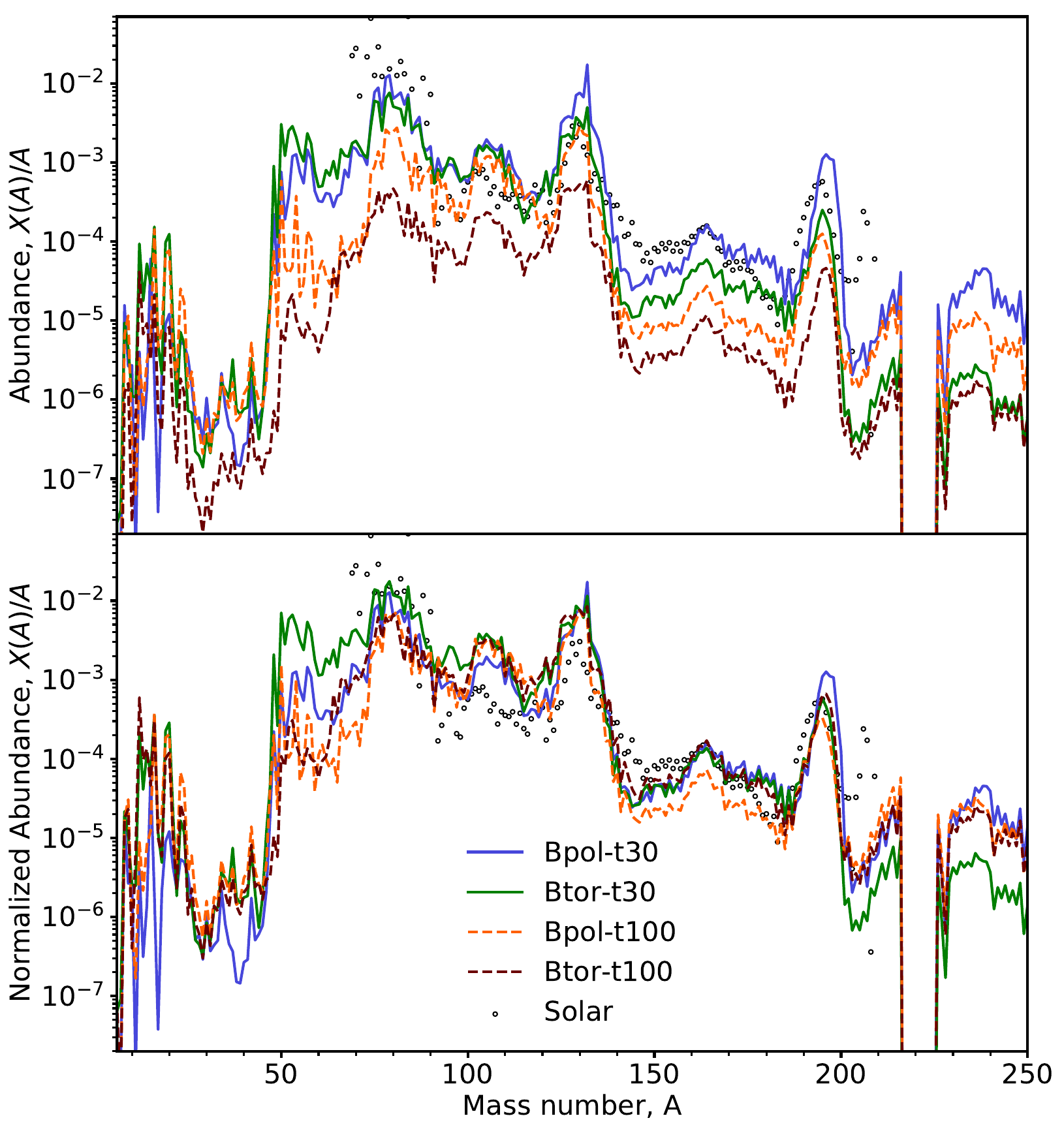}
\caption{\emph{Top:} Abundances at 30 years computed with the nucleosynthesis code \texttt{SkyNet} using
the trajectories of all unbound tracer particles past 1000\,km in the simulation. Shown in
open circles are the solar r-process abundances from \citet{Goriely1999}, scaled
such that the abundance at the second peak (A=130) matches that of our
\emph{Bpol-t30} model.
\emph{Bottom:} Same as top, except that \textit{all} models are scaled such that they match the abundances of \emph{Bpol-t30} at A=130, to illustrate differences in abundance pattern assuming that the ejected mass is the same.
} 
\label{fig:skynet} 
\end{figure}

\subsection{Tracer Particles and Nucleosynthesis}

The r-process abundance patterns at $t=30$\,yr 
for all of our simulations are shown in Figure~\ref{fig:skynet}.   
All models broadly follow the solar r-process abundance pattern. 
Both increasing the HMNS lifetime and initializing with a toroidal field geometry result in less mass ejection on timescales shorter than those required for weak interactions to raise the electron fraction above the critical value of $Y_e \gtrsim 0.25$. As a result of this, we see a drop in abundances with $A > 130$ in these models compared to the \emph{Bpol-t30} model, qualitatively consistent with other studies.

The lifetime of the HMNS has the largest impact on the abundance distribution, with a decrease of almost 100 times from model \emph{Bpol-t30} to \emph{Bpol-t100} in  these elements. However, the significant decrease in initially purely magnetic driven outflows (e.g., between \emph{Bpol-t30} and \emph{Btor-t30}) also causes a drop by almost half an order of magnitude. This is consistent with the
expectations from the distribution of our particles in $s_\mathrm{k_B}-Y_e-v$ space. 

Overall, we find that all 4 models produce the 3 process peaks.
By normalizing \textit{all} the abundance patterns to the second peak of \emph{Bpol-t30}, we find that the relative ratios of light to heavy r-process elements are very similar between models. The \emph{Bpol-t100} model tends to underproduce the rare-earth peak, which could be due to the increased high entropy ($s \gtrsim 100 k_\mathrm{B}/\mathrm{baryon}$) neutrino-driven winds during $30\,\mathrm{ms}\leq t \leq 100\,\mathrm{ms}$ that makes lighter seed nuclei for the r-process to build on \citep{Lippuner2015}. We speculate that if more thermally driven outflows with lower entropy and $Y_e \sim 0.3$ were captured by running the simulation longer, this discrepancy may vanish as the fraction of ejecta from neutrino driven winds decreases. We also see a relative underproduction of actinides and overproduction of lighter $(A \lesssim 100)$ r-process elements in \emph{Btor-t30}. This is consistent with the additional high $Y_e$ material ejected during HMNS collapse, which makes a substantial contribution to the total ejected mass. 

\begin{table*} 
\caption{Mass ejection from all models. Columns show, from left to right, the
 model name, maximum simulation time, total unbound mass ejected at
$r_\mathrm{out} = 1000\,\mathrm{km}$ using the
Bernoulli criterion, mass ejected that is composed of HMNS material, 
$\dot{M}_\mathrm{out}$-weighted average electron fraction and radial velocity, as
well as unbound ejected mass, average electron fraction, and radial velocity
broken down by electron fraction (superscript blue 
lanthanide-poor: $Y_e \geq 0.25$, red lanthanide-rich: $Y_e < 0.25$).}
\centering
\resizebox{\textwidth}{!}{%
\begin{tabular}{l|ccccccccccc} 
Model & $t_\mathrm{max}$ 
& $M_\mathrm{out}               $ & $M_\mathrm{out}^\mathrm{hmns} $ & $\la Y_e \ra               $ & $\la v_r\ra $  
& $M^\mathrm{blue}_\mathrm{out} $ & $ \la Y_e^\mathrm{blue}\ra    $ & $\la v^\mathrm{blue}_r \ra $  
& $M^\mathrm{red}_\mathrm{out}  $ & $ \la Y_e^\mathrm{red} \ra    $ & $\la v^\mathrm{red}_r  \ra $  \\
& (s) 
& $ (10^{-2} M_\odot)           $ & $(10^{-2} M_\odot) $ &   & $ (c) $                       
& $ (10^{-2} M_\odot)           $ &                          & $ (c) $                       
& $ (10^{-2} M_\odot)           $ &                          & $ (c) $  \\ 
\hline\hline
Bpol-t30  & 1.31 & 6.309 & 0.012 & 0.235 & 0.057 & 2.073 & 0.295 & 0.125 & 4.236 & 0.206 & 0.024 \\
Btor-t30  & 2.00 & 5.992 & 0.026 & 0.226 & 0.035 & 1.494 & 0.305 & 0.091 & 4.498 & 0.200 & 0.017 \\
Bpol-t100 & 0.75 & 4.790 & 0.012 & 0.223 & 0.059 & 1.619 & 0.294 & 0.120 & 3.171 & 0.186 & 0.029 \\
Btor-t100 & 0.70 & 1.886 & 0.021 & 0.213 & 0.050 & 0.510 & 0.317 & 0.109 & 1.376 & 0.175 & 0.028 \\
\hline
\end{tabular}%
}
\label{tab:outflowresults} \end{table*}

\section{Comparison to previous work} \label{sec:comparison}
\subsection{3D Simulations}
\citet{Kiuchi2022} carry out simulations in full GRMHD of a magnetized NS-NS merger, including neutrino leakage and absorption, and using the SFHO EOS. They run the simulation until $\sim1.1\,\mathrm{s}$ post merger, with the HMNS surviving for 17 ms, and report mostly red post-merger outflows with broad range of electron fractions peaking at $Y_e \sim 0.24$ and traveling at $v \lesssim 0.15c$. The material with $Y_e \gtrsim 0.25$ is ejected via turbulent angular momentum transport from MRI operating in the disk after HMNS collapse, which travels too slow to power the blue kilonova from GW170817, consistent with other MHD simulations of BH-torus ejecta \citep{Hayashi2021,FahlmanFernandez2022,Curtis2023}. They note a lack of magnetic ``tower" structure which drives outflows in the polar regions. Our \emph{Btor-t30} model most closely matches the post-merger state found in their work, and our outflows show broad agreement in the electron fraction and entropies of the ejecta, as well as a lack of ``tower" outflows. We do not however find a sharp cutoff in post-merger mass ejection with velocities $v > 0.15c$, but rather a gradual fall off. This could be due to the increased lifetime of our HMNS, as well as the lack of special relativistic effects limiting the velocity of our outflows. 

The simulations of \citet{Combi2023Jet} are also of a GRMHD neutron star binary with neutrino leakage and absorption, using the APR EOS. Their HMNS survives $\sim60\,\mathrm{ms}$ (the duration of the simulation), and ejects $\sim 10^{-2}\,M_\odot$ of ejecta, travelling with $v \gtrsim 0.1c$ mainly through disk winds. Through irradiation from the HMNS, the vast majority of this ejecta has $\langle Y_e \rangle \sim 0.3$. In our poloidal setups we also find sustained mass ejection rates of $\sim 0.1\,M_\odot\,s^{-1}$. They find about $5\%$ of their outflows are generated from HMNS magnetized ``tower" outflows, consistent with our simulations where we find only $\sim 10^{-4}\,M_\odot$ ($\sim 1\%$) come from magnetized outflows from the HMNS. They also find similar amounts of matter, $\sim 10^{-3}\,M_\odot$, moving with $v > 0.25c$ and $Y_e > 0.25$, which is consistent with our simulations. 

Interestingly, they find that this result is consistent with the early blue kilonova of GW170817 through a simple kilonova model \citep{Combi2023}. This is consistent with the recent kilonova models of \citet{Bulla2023} and \citet{Ristic2023}, which show that massive ($\sim 10^{-2}\,M_\odot$), blue ($Y_e > 0.25$), disk winds with $v \sim 0.05c$ are sufficient to power the early blue kilonova at times $\lesssim$5 days. This suggests that our very similar outflows would also be able to power the blue kilonova, although this cannot be confirmed without self-consistent modelling.

\citet{deHaas2022} examine the effects of magnetic field strength and geometry within the HMNS on the outflows mapped from a post-merger system, using 3D GRMHD neutrino-leakage/absorption simulations with the LS220 EOS. They find that a magnetar strength poloidal magnetic field in the HMNS ($B\sim 10^{15}\,\mathrm{G}$) is capable of ejecting $\sim 10^{-3}\,M_\odot$ of ejecta travelling with a wide range of velocities, $0.05c < v <0.6c$, and $Y_e \gtrsim 0.25$. They find decreasing the field strength by an order of magnitude decreases the ejecta
to $\sim 10^{-4}\,M_\odot$, while also lowering the maximum velocity of the ejecta to $\sim 0.2c$. As well, they show changes in the imposed field geometry have similar effects. We find similar distributions of ejecta in velocity and electron fraction in our simulations, with mass ejection rates similar in our poloidal runs, despite in our simulations imposing a weaker ($10^{14}\,\mathrm{G}$) initially toroidal field within the HMNS, but this changes quickly $(\lesssim 20\,\mathrm{ms}$) through the dynamics inside the HMNS. The field acquires a large poloidal component that peaks at values of $3\times 10^{16}\,\mathrm{G}$, and in the case of the 100\,ms HMNS saturates at $8\times 10^{16}\,\mathrm{G}$. Additionally, they find changing magnetic field geometry to be primed for more stress driven outflow results in less $A>130$ element nucleosynthesis, consistent with our findings.

\citet{Curtis2023HMNS} and \citet{Mosta2020} perform 3D GRMHD simulations using a two moment (M1) and
neutrino leakage scheme, respectively, the LS220 EOS, and the same HMNS remnant that survives for 12 ms after mapping in from a hydrodynamic merger simulation at 17\,ms  post-merger with an added $B\sim 10^{15}\,\mathrm{G}$ poloidal field. Both studies find $\sim3\times 10^{-3}\,M_\odot$ worth of material ejected and velocities peaking at $0.15c$, with a significant tail up to $0.5c$. They highlight the difference in using an M1 vs leakage scheme for handling neutrinos, as the more advanced M1 scheme shifts the peak electron fraction of the ejecta up from distribution around $Y_e \sim 0.25$ to one peaking at $Y_e \sim 0.35-0.45$. Our results are similar in both velocity and electron fraction to the outflows from their system, especially to the leakage results of \citet{Mosta2020}, with a mass ejection rate very similar to that of our \emph{Bpol-t30} and \emph{Bpol-t100} models.

\subsection{2D Simulations}
Studies that use axissymmetric simulations find that neutrino driven winds from
the HMNS can reach velocities of $v \sim 0.15c$, although the exact velocity, ejecta mass, and composition depends on the lifetime of the HMNS, prescription used for neutrino radiation, as well as the handling of angular momentum transport. Higher velocities are possible, but they come with increased irradiation of the ejecta, making a simultaneous match to the blue and red KN difficult (e.g.,\citealt{FahlmanFernandez2018,Lippuner2017,Nedora2021,Fujibayashi2023}). 

However, the recent hydrodynamic simulations of \citet{Just2023}, mapped in from a merger simulation, include a more advanced energy-dependent neutrino leakage scheme and utilize the SFHo EOS, as well as varying remnant survival times in pseudo Newtonian potential. They
find neutrino driven winds with a mass of $\sim 10^{-2}\, M_\odot$ and velocity $\langle v_\mathrm{ej}\rangle \sim 0.2c$ for their remnants which survive for a comparable amount of time, $\sim 100\,\mathrm{ms}$. The mass ejection rates are broadly similar to our simulations $\sim 10^{-2} - 10^{-1}\,M_\odot\,\mathrm{s}^{-1}$, and due to the ejection mechanism, these tend to be high electron fraction $Y_e > 0.25$. They produce few elements with $A > 130$, most consistent with our \emph{Btor-t100} run, which in our case produces the most dominant neutrino driven wind. 

As well, \citet{Shibata2021} perform unique 2D resistive GRMHD simulations, with a mean field prescription to prevent the damping of the magnetic dynamo in axisymmetry. Their
simulations start from prescribing a toroidal magnetic field onto the
outcome of a GRHD merger simulation using the DD2 EOS, similar to our idealized initial
conditions. The setup of their low resistivity
simulations are most comparable to our ideal MHD treatment, although their remnant has a different
rotational profile, and survives the for the duration of the simulation. They find
ejecta masses of $\sim0.1\,M_\odot$ that plateau at $\sim 500\,\mathrm{ms}$, with average velocities of $0.5c$. The velocity and electron fraction of the ejecta is comparable to our
\texttt{Btor-t30} run, although we find more low electron fraction ejecta and
less mass ejected, by an order of magnitude.      

\subsection{Discussion}

In general, our results are in agreement with those of the literature. They tend to span a broader range of electron fraction than is found by other studies in the literature, in particular with a larger component of material ejected with $Y_e < 0.2$. We speculate this is due to the simplicity of our leakage scheme in comparison to the more advanced energy dependent leakage, M1, or MCMC schemes (See \secref{sec:AppA}) in combination with the idealized initial conditions for the torus, which tends to eject material in fast, magnetic stress driven outflows that can escape neutrino interactions. 

Our HMNS itself has a lower mass ejection rate than others found in the literature. This is also likely partially due to the neutrino scheme, which yields less efficient heating of matter surrounding the HMNS, with implications for the neutrino driven winds. Additionally, the importance of the magnetic field configuration and resolution within and around the HMNS likely plays a large role. Our toroidal field embedded in the HMNS ejects mass similar to that of \citet{Kiuchi2022}, but ejects an order of magnitude less mass than poloidal field configurations (e.g., \citealt{Combi2023Jet,deHaas2022, Curtis2023HMNS}). We do not resolve the most unstable wavelength of the MRI inside our HMNS, meaning that MRI driven ejection is not captured. Finally, we note that changing the rotation profile of the HMNS to match those of merger simulations could result in additional mechanical-oscillation powered outflows as angular momentum is transported in the HMNS. 

\section{Conclusions} 
\label{sec:conclusion}
We have performed $\sim 1\,\mathrm{s}$ long 3D MHD simulations of a an idealized post-merger system consisting of a 2.65$M_\odot$ HMNS and $0.1\,M_\odot$ torus. We utilize Newtonian  self-gravity, the hot APR EOS, a leakage/absorption scheme to handle neutrino interactions, and a pseudo-Newtonian potential after
BH formation. Motivated by the sensitivity of the HMNS collapse time to physical processes, and necessitated by our use of Newtonian gravity, we use two parameterized HMNS collapse times of 30\,ms and 100\,ms to determine the effects of a HMNS as a central remnant. To evaluate the effects of the initial magnetic field geometry we utilize either a toroidal or poloidal magnetic field threaded through the torus.

The outflows are similar to those produced by idealized BH-disk setups, with a broad distribution of electron fraction and velocities. The HMNS itself tends to drive additional fast ($v\gtrsim 0.2c$) high electron fraction outflow ($Y_e > 0.3$) from the torus while it survives, due to oscillations in the remnant and energy from neutrino irradiation. Upon collapse, accretion onto the BH drives additional outflows, and slower ($v < 0.1c$), redder ($Y_e \sim 0.2$) MRI driven outflows begin to dominate the total mass ejection. 

We find that in all cases, a shock wave is launched upon collapse of the HMNS as the torus and newly formed BH settle into an accreting state. The creation of a rarefraction wave has been seen in previous 2D hydrodynamic simulations \citep{FahlmanFernandez2018}, but in the magnetized 3D case we find it drives significant outflows, especially noticeable in the short-lived HMNS. The launching of a magnetized shock by supramassive NS collapse has been explored in baryon-free environments in the context of powering fast radio bursts (the ``blitzar'' mechanism, \citealt{Most2018}), but it is unknown whether this shock would drive mass outflows in a baryon polluted system. Whether this effect is due to our instantaneous collapse in Newtonian gravity, idealized initial conditions, or is a real effect, remains to be seen. 

Neutrino driven winds from a longer lived ($t > 30\,\mathrm{ms}$) HMNS tend to be suppressed by the accretion of matter onto the HMNS, which cools efficiently. Due to accretion from the torus and oscillations of the HMNS, we find a lack of high velocity ``tower" ejecta from the polar regions, which tend to be too dense to eject large amounts of matter moving at relativistic velocities. 

Nucleosynthesis in the outflows tends to predict a robust r-process up to the second peak ($A\sim130$), with order of magnitude variations past $A\sim130$ caused by changing HMNS lifetime or magnetic field geometries. 

Across all models, we find $M\sim 10^{-3}M_\odot$ of ejecta with $Y_e > 0.25$ and $v > 0.25c$. Whether or not this can completely account for the blue kilonova of GW170817 is model-dependent, as simple 2-component models require an order of magnitude more blue, fast ejecta. However, multidimensional models with more realistic opacities, thermalization, heating rates and viewing angle dependence find that a massive ($10^{-2} \,M_\odot$), slower ($v\sim 0.1c$) wind is all that is required. If the latter is the case, then our outflows are in principle capable of powering the blue kilonova.

The main limitations of our simulations is the use of an approximate neutrino scheme and Newtonian gravity. In testing of our neutrino scheme (see \secref{sec:AppA}), we find results similar to that of other leakage comparisons, with an under prediction of neutrino heating resulting in slower, lower $Y_e$ outflows as compared to more robust Monte Carlo or M1 schemes (e.g., \citealt{ILEAS2019,Radice2021,Curtis2023HMNS}). 
For a similar mass HMNS in GR, we estimate that the HMNS could be up to 1.5 times as compact as our Newtonian HMNS upon merger. This reduces the amount of neutrino emission in the Newtonian HMNS compared to a general relativistic one, and results in the same under predictions of neutrino absorption in the torus, an effect which has been documented in the context of core collapse supernova simulations (e.g.  \citealt{Liebendorfer2001,Marek2006,Mueller2012b,OConnor2018,Mezzacappa2023}). As well, although we find a lower density magnetized funnel in our simulations, it is unlikely that we are able to correctly model jet formation in our simulations. Inclusion of special relativistic effects to the MHD equations results in corrections that are of order $\sim v/c$, so we estimate uncertainties in the fast tail of our ejecta of $\sim 50\%$, but the majority of outflows sit at v < 0.1c, which results in a $10\%$ inherent uncertainty.

\section*{Acknowledgements}
We thank Coleman Dean and Suhasini S. Rao for helpful suggestions and discussions.
SF and RF are supported by the Natural Sciences and Engineering Research
Council of Canada (NSERC) through Discovery Grant RGPIN-2022-03463 and SM by NSERC Discovery Grant RGPIN-2019-06077.  
The software used in this work
was in part developed by the U.S. Department of Energy (DOE) 
NNSA-ASC OASCR Flash Center at the University
of Chicago.  
Data visualization was done in part using \texttt{VisIt} \citep{VisIt}, which is supported
by DOE with funding from the Advanced Simulation and Computing Program
and the Scientific Discovery through Advanced Computing Program.
This research was enabled in part by support
provided by WestGrid (www.westgrid.ca), the Shared Hierarchical
Academic Research Computing Network (SHARCNET, www.sharcnet.ca),
Calcul Qu\'ebec (www.calculquebec.ca), and Compute Canada (www.computecanada.ca).
Computations were performed on the \emph{Niagara} supercomputer at the SciNet
HPC Consortium \citep{SciNet,Niagara}. SciNet is funded by the Canada
Foundation for Innovation, the Government of Ontario (Ontario Research Fund - Research Excellence),
and by the University of Toronto.

\section*{Data Availability}

The data underlying this article will be shared on reasonable request to the corresponding author.
\bibliography{ms}
\bibliographystyle{mnras}
\appendix
\section{HMNS Neutrino Leakage Scheme} \label{sec:AppA}
In this appendix we detail the implementation of additions to our neutrino
scheme to include the effects of HMNS irradiation in the domain, in particular the effects of neutrino heating on the torus. Full details of the scheme are outlined in \citet{FahlmanFernandez2022}. The end goal of the implementation is scheme accurate to within an order of magnitude with limited computation costs. It is well known that neutrino leakage schemes tend to underpredict the lepton number change and energetics in merger simulations (e.g., \citealt{Foucart2020MC, Radice2021, Curtis2023HMNS}), and as such we do not attempt to create a totally quantitatively accurate scheme. 

\subsection{Neutrino Implementation}

The HMNS is a source of neutrino emission, so we change our
leakage-lightbulb scheme to have an additional component of heating from the
HMNS applied to the domain. We extend the implementation of
\citet{MF14} and \citet{Lippuner2017} for heating fluid elements outside the neutrinosphere,
which requires the location of the neutrinosphere, as well as the
temperature and luminosity of the neutrino emission.

\begin{figure}
\begin{picture}(400,400)
\put(0,0){\includegraphics[width=\columnwidth]{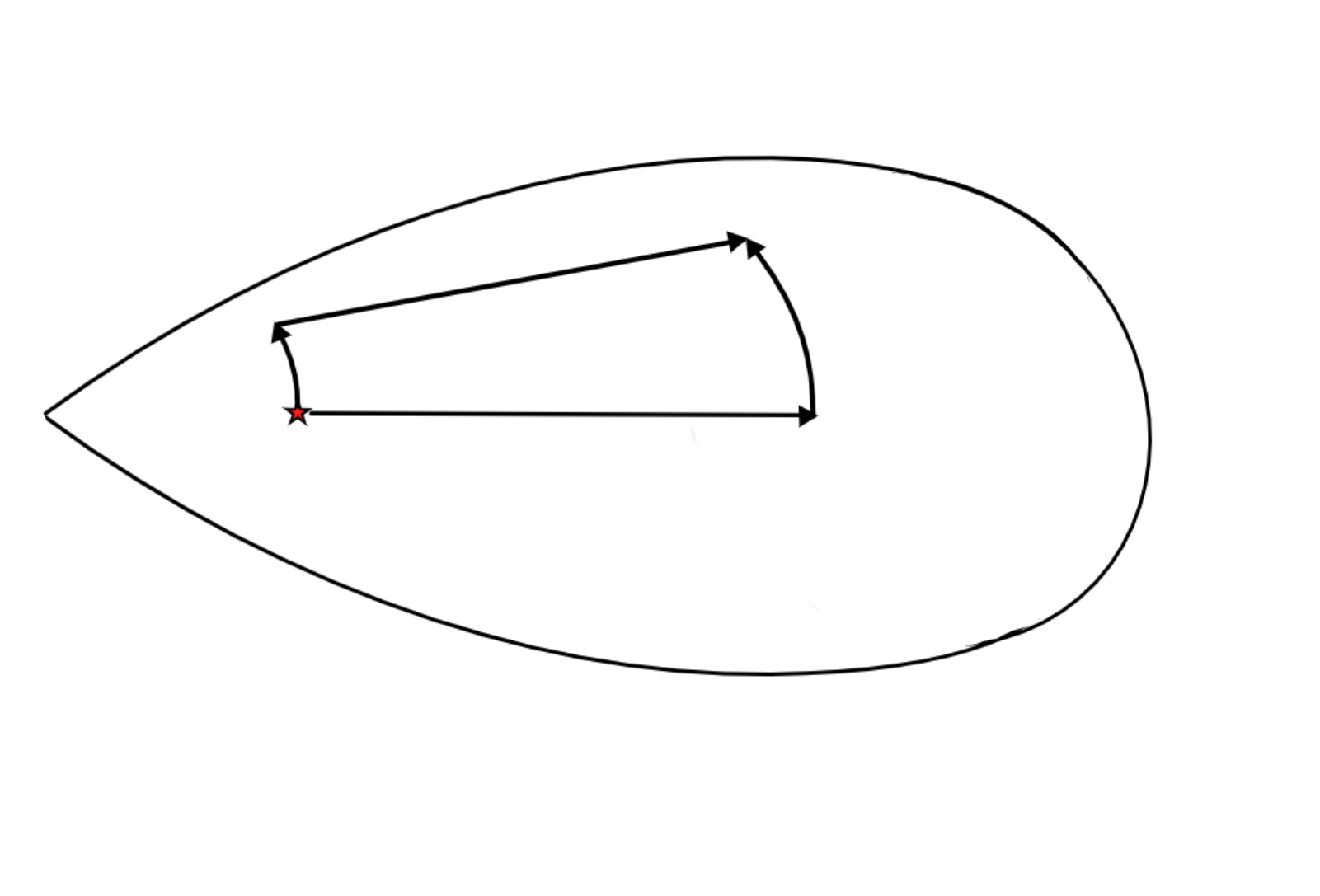}}
\put(40,95){\large $\tau_3$}
\put(150,103){\large $\tau_4$}
\put(100,80){\large $\tau_2$}
\put(100,120){\large $\tau_1$}
\put(55,80){\large $R_\mathrm{em}$}
\put(143,122){\large $P(r,\theta)$}
\end{picture}
\caption{Integrated line of sight optical depths used in determining the
reduction in self-irradiation heating from torus emission. The torus is taken to
be a lightbulb which emits from two points slightly above and below the the
midplane at a radius $R_\mathrm{rm}$, to each point in the domain, $P(r,\theta)$ }
\label{fig:TorusAbsSchematic}
\end{figure} 

The explicit heating and cooling terms are given in the appendices of
\citet{FM13} and \citet{MF14}. What requires modification is the total neutrino luminosity
input from the HMNS, which is then normalized by the blackbody
radiation of the neutrinos at the neutrinosphere, and a radial component to
account for the flux hitting the cell,
\begin{align}\label{eq:normLnu}
  \bar L_{\nui} = \frac{L_{\nui}}{4\pi R_{ns,\nui}^2 \sigma_{SB} T_{ns,\nui}^4}.
\end{align} 
In previous simulations, the neutrino luminosity $L_\nui$, neutrinosphere temperature, $T_{ns,\nui}$, and neutrinosphere radius $R_{ns,\nu}$ were
user defined parameters, and we now make them self-consistent. We find
the total luminosity of the HMNS by summing up the individual emissivities of
each cell within the HMNS. The neutrinosphere radius is not assumed to be
spherically symmetric, and is allowed to change with $\theta$ and $\phi$. It is
determined as the first location radially outwards at which the optical depth is
less than or equal to 2/3,
\begin{align}
  R_{ns,\nui}(\theta,\phi) = R(\tau_\nui \leq 2/3 ).
\end{align}
The radial optical depth for each species is calculated along each direction using the numerical summation $\tau_\nui = \sum \kappa_\nui(r) \d r$, where we utilize the analytic expressions for $\kappa$ from \citet{Ruffert1996}, which include the effects of  energy averaged scattering and absorption processes onto neutrons and protons.
The neutrino temperature is then determined from the average neutrino energy in
the neutrinosphere
\begin{align}\label{eq:neutrinoT}
  \langle \epsilon_{ns,\nui} \rangle &=
\sum{Q_\nui^\mathrm{eff}(R_{ns})}/\sum{R_\nui^\mathrm{eff}(R_{ns})}, \\
  \langle T_{ns,\nui} \rangle &= \langle \epsilon_{ns,\nui} \rangle
\frac{\mathcal{F}_4(0)}{\mathcal{F}_5(0)},
\end{align}
where $Q_\nui^\mathrm{eff}$ and $R_\nui^\mathrm{eff}$ are the effective energy and lepton number change for each neutrino species, the explicit forms for which are defined in \citet{Fernandez2019}, eq (3) and (4).  The factors of the Fermi integrals, $\mathcal{F}_n(0)$ come from expanding
the explicit forms of $Q^\mathrm{eff}_\nui$ and $R^\mathrm{eff}_\nui$. 

The neutrino radiation is then further attenuated by a factor of
$e^{-2\tau_\nui}$, where $\tau_\nui$ is the radially integrated optical depth
along the line of sight from the cell to the HMNS.  

Within the neutrinosphere, the irradiation is identical, except that it relies
on the spherically symmetric enclosed luminosity and temperature at each radial
cell. These enclosed quantities are then used in equations~
(\ref{eq:normLnu}-\ref{eq:neutrinoT}) to calculate the normalized luminosity used in
the heating and lepton number change rates. In practice, however, the
luminosity within the HMNS is dampened to negligible rates by the exponential of
the massive opacities ($\tau_\nui \sim 10^4$), which drops steeply across one
single cell at the HMNS neutrinosphere with our current resolution.

Additionally, we improve the lightbulb scheme for emission from the torus by
integrating $\tau_\nui$ for absorption from neutrinos
emitted by the torus. While this is a subdominant effect compared to both
cooling from the torus and irradiation from the HMNS, the previous scheme of \citet{FahlmanFernandez2022} has
problems with high density tori, especially after they experience shocks from
the HMNS oscillation. These shocks can artificially increase $\tau_\nui$ in the outer regions of the torus, as the previous scheme used the value of $\tau_\nui$ at
the density maximum for the luminosity reduction in the entire torus. 

We develop a simple new scheme which relies on integration of $\tau_\nui$ in the
radial and polar directions. First, we determine the 4 optical depths shown in Figure~\ref{fig:TorusAbsSchematic}. The first is
along the radial line of site from the emission maximum to each point
$(\tau_1)$, the radial optical depth along the equator to the radius at which the
point lies ($\tau_2)$, the angular optical depth from the torus equator to the emission
maximum ($\tau_3$) and to the point ($\tau_4$). The optical depth is then taken
to be the maximum of the two possible path lengths
\begin{equation}
\tau = \max(\sqrt{\tau_1^2 + \tau_3^2}), \sqrt{\tau_2^2+\tau_4^2}).
\end{equation}
to avoid issues with points around the HMNS where one of the path can have unphysically
small optical depths. This is similar to other neutrino leakage schemes (e.g., \citealt{Ruffert1996, Neilsen2014,Siegel2018,Werneck2023}).

\begin{figure}
\includegraphics[width=\columnwidth]{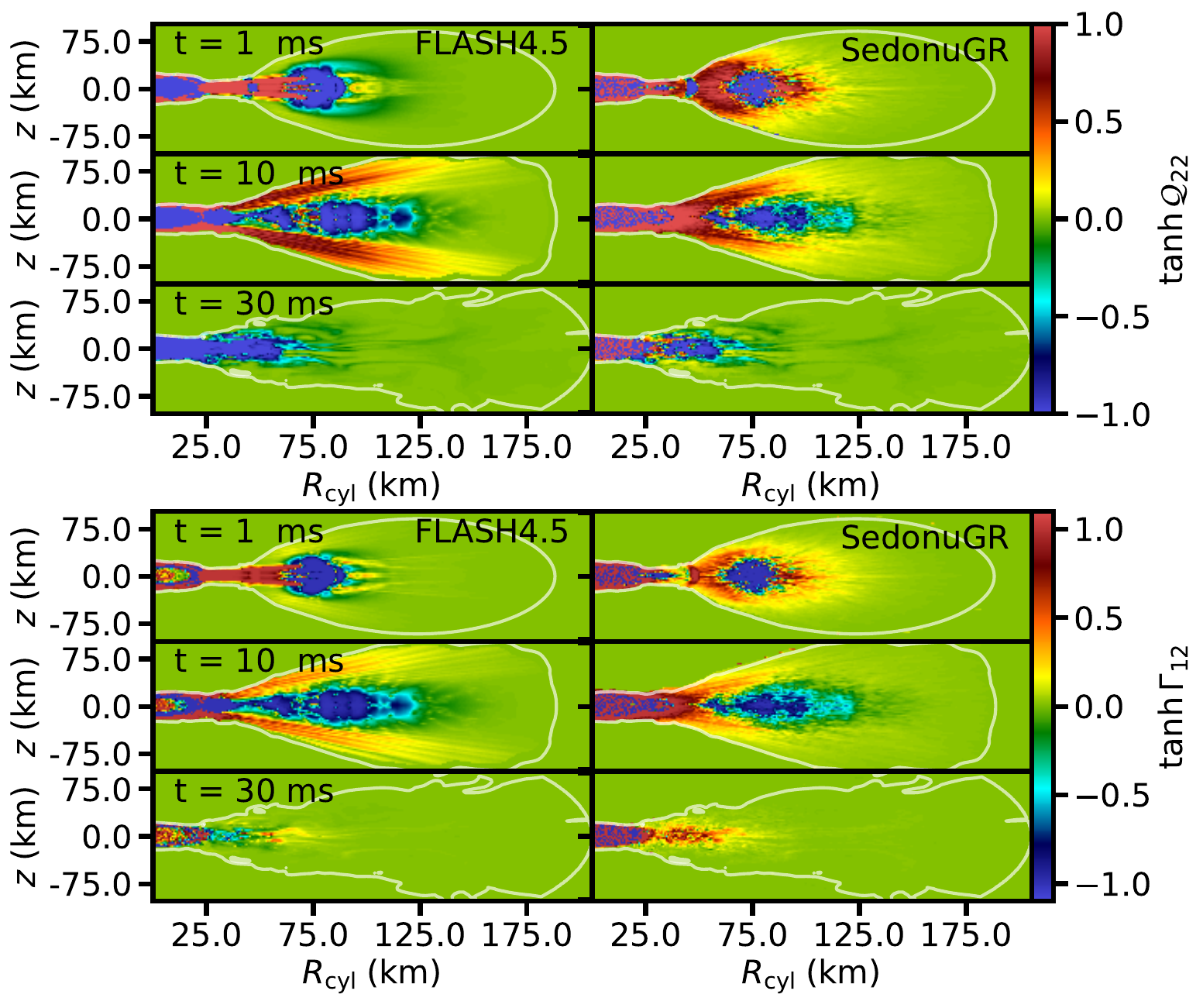}
\caption{Comparison of neutrino source terms for a $0.1\msun$ torus and $2.65\msun$ HMNS at $t=0,1,30$ ms in 
\texttt{FLASH4.5} (\textit{Left})  to \texttt{SedonuGR} (\emph{Right}). The top 6 panels compare rate of energy change per unit mass, and the bottom 6 panels show rate of lepton number change per baryon. The background fluid snapshot are taken as 2D slices at $\phi=0$ from the \texttt{FLASH4.5} domain at the specified time, with the $10^9 \mathrm{g\,cm}^{-3}$ white contours delimiting the approximate edge of the torus, where the neutrino source terms are set to 0. The lightbulb/leakage scheme tends to underestimate the heating and lepton change at all times, leading to less energetic, lower $Y_e$ neutrino driven outflows than in an MC scheme.}
\label{fig:HMNSNeutrinoComparison}
\end{figure}

\begin{figure}
\includegraphics[width=\columnwidth]{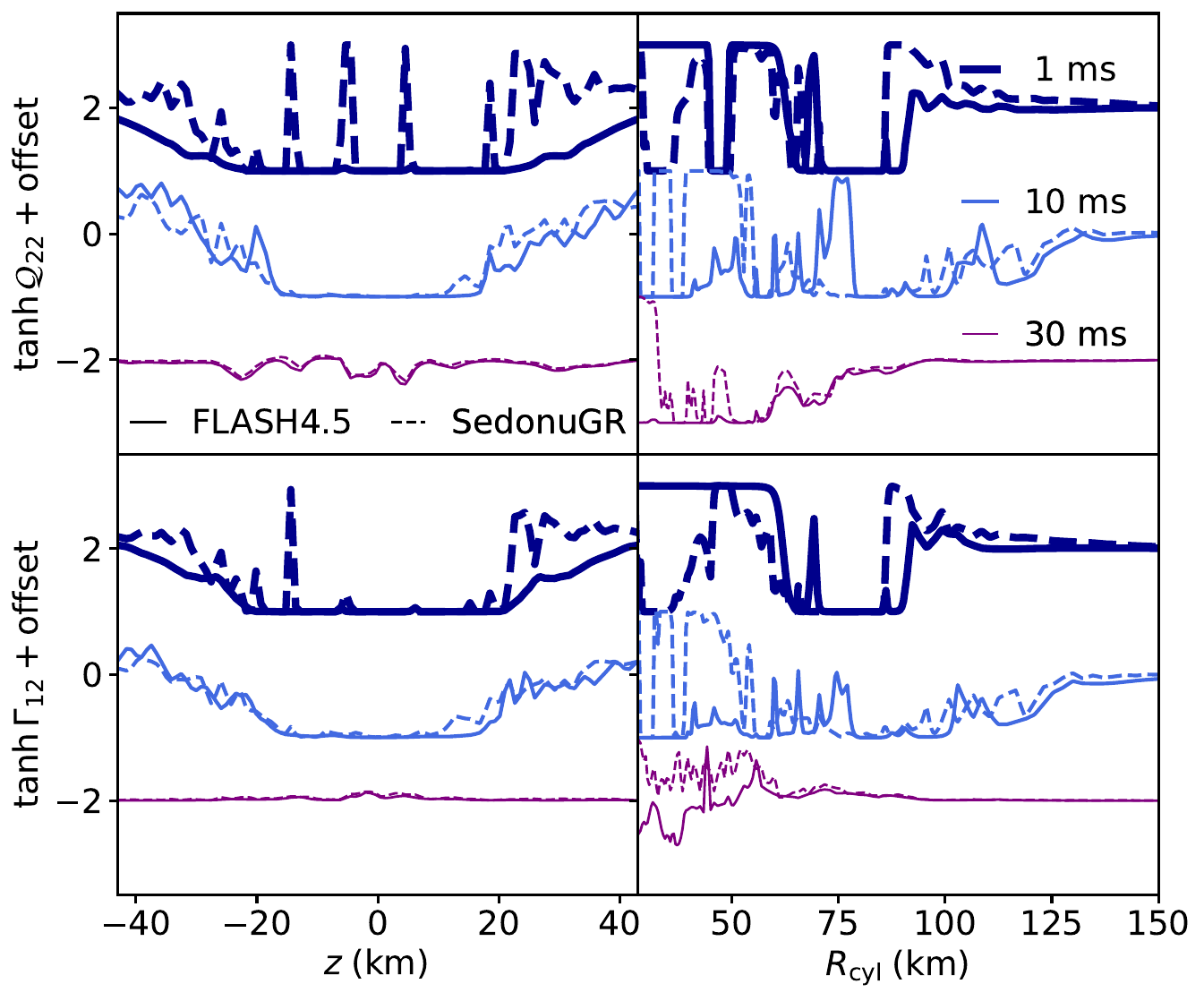}
\caption{Comparison of neutrino source terms for a $0.1\msun$ torus and $2.65\msun$ HMNS at $t=0,1,30$ ms in \emph{Top}: Rate of energy change per unit mass, and \emph{Bottom}: Rate of lepton number change per baryon. The absolute values of the source terms for $t=1,30$\,ms have been offset by a constant ($\pm 2) $ for each time for visual clarity. The background fluid snapshots are from the \texttt{FLASH4.5} domain at the specified time. \emph{Left:} Vertical slices at $\phi=0$ and $r=r_\mathrm{circ}$ (through the torus density maximum). \emph{Right:} Radial slices at $\phi=0$ and $\theta=90$ (through the equator). The neutrino source terms from \texttt{FLASH4.5} are shown as solid lines, while the source terms from \texttt{SedonuGR} are dashed lines. Changing color and line thickness denote different time snapshots.}
\label{fig:HMNSNeutrinoSlices}
\end{figure}

We compare snapshots of the net energy and lepton number change source terms from our scheme to
the time stationary Monte-Carlo (MC) neutrino
transport code \texttt{SedonuGR} \citep{Richers2015,Richers2017} in Figure~\ref{fig:HMNSNeutrinoComparison} and Figure~\ref{fig:HMNSNeutrinoSlices}.
\texttt{SedonuGR} utilizes an MC algorithm accounting for emission, absorption, and scattering of
neutrinos for a given fluid background and EOS to return the local energy and
lepton number change rates for a fluid parcel. MC schemes are among the most accurate in the literature (e.g., \citealt{Richers2015,Ryan2015,Richers2017,Miller2019,Foucart2020MC}), so the source terms returned by
\texttt{SedonuGR} are a good benchmark that we can compare our more approximate
leakage/lightbulb scheme to. Snapshots from 2D slices of our spherical domain at $t=0,1$ and $30$\,ms are taken as comparison points. The effects of neutrinos start to become negligible due to decreased
emission and increased transparency as the torus dissipates, and for this reason, we focus our comparison at earlier timesteps, with earlier agreement being much more impactful.

The neutrino scheme recreates the overall energy and lepton number change rates to within a factor of a few, following a similar spatial distribution. As the scheme progresses, cooling of material is overestimated by an order of magnitude, especially in the midplane of the torus and close to the HMNS. We show this both with 2D colormaps, as well as a numerical comparison of the schemes by taking a radial slice through the domain at the equator, and a vertical slice at the density maximum of the torus. While it is difficult to extrapolate the effects of the discrepancies on our outflows, we can speculate that increased cooling near the midplane of the HMNS may result in less neutron rich material being ejected by magnetically driven outflows, which is also found by \citet{Curtis2023HMNS} in their leakage scheme comparison. However, the majority of matter affected is also likely to be accreted upon HMNS collapse, as its small angular momentum causes it to circularize at a radius smaller than the ISCO upon collapse to BH.
 
\bsp
\label{lastpage}
\end{document}